\documentclass[prl,twocolumn,superscriptaddress]{revtex4-2}
\usepackage{graphicx}
\usepackage{amsmath,amssymb}
\usepackage{natbib}
\usepackage[colorlinks,linkcolor=blue,anchorcolor=blue,urlcolor=blue,citecolor=blue]{hyperref}
\usepackage[all]{hypcap}
\usepackage{mathrsfs}
\usepackage{bm}
\usepackage{bbm}
\usepackage{dsfont}
\usepackage[utf8]{inputenc}
\usepackage[english]{babel}
\usepackage[T1]{fontenc}
\usepackage{newtxtext}

\begin{document}

\title{Strong angular momentum optomechanical coupling for macroscopic quantum control}
\author{Yuan Liu}
\author{Yaoming Chu}
\author{Shaoliang Zhang}
\email{shaoliang@hust.edu.cn}
\author{Jianming Cai}
\email{jianmingcai@hust.edu.cn}
\affiliation{School of Physics, Hubei Key Laboratory of Gravitation and Quantum Physics, Institute for Quantum Science and Engineering, International Joint Laboratory on Quantum Sensing and Quantum Metrology,  Huazhong University of Science and Technology, Wuhan 430074, China}
\affiliation{Wuhan National Laboratory for Optoelectronics, Huazhong University of Science and Technology, Wuhan 430074, China}

\date{\today}

\begin{abstract}
Optomechanical systems offer unique opportunities to explore macroscopic quantum state and related fundamental problems in quantum mechanics. Here, we propose a quantum optomechanical system involving exchange interaction between spin angular momentum of light and a torsional oscillator. We demonstrate that this system allows coherent control of the torsional quantum state of a torsional oscillator on the single photon level, which facilitates efficient cooling and squeezing of the torsional oscillator. Furthermore, the torsional oscillator with a macroscopic length scale can be prepared in Schrödinger cat-like state. Our work provides a platform to verify the validity of quantum mechanics in macroscopic systems on the micrometer and even centimeter scale.
\end{abstract}

\maketitle

{\it Introduction.---} Optomechanics is currently under intense research focus \cite{marquardt2007quantum,schliesser2006radiation,wilson2007theory,hoff2016measurement,li2018generation,teh2018creation,Akram_2013,teh2018creation,PhysRevLett.99.093901,PhysRevLett.97.243905}, aiming for coherent manipulation of the quantum states of macroscopic objects \cite{marquardt2007quantum,schliesser2006radiation,wilson2007theory,hoff2016measurement,li2018generation,teh2018creation}. As an versatile platform to investigate fundamental problems in quantum mechanics, quantum optomechanics provides a route towards generation of macroscopic quantum states \cite{PhysRevLett.99.093901,PhysRevLett.97.243905, Akram_2013,teh2018creation,PhysRevLett.126.033601,PhysRevA.98.011801}. It is remarkable that macroscopic quantum states, such as Schrödinger cat states \cite{1935} (namely coherent superposition of two macroscopically distinct states), phonon addition and subtraction states \cite{PhysRevLett.126.033601,PhysRevA.98.011801}, are of great significance in exploring the boundary of quantum theory \cite{RevModPhys.90.025004} and the validity of collapse models \cite{RevModPhys.85.471}.
In most optomechanical systems, the optomechanical interaction between light and matter is induced by the linear momentum exchange interaction. However, the angular momentum of light can also be utilized to contribute the exchange interaction between light and matter \cite{PhysRevA.66.063402,PhysRevA.54.1593,PhysRevA.75.063409,Friese1998,Paterson912}.
This opens the door to new possibilities in the engineering of optomechanical system. For example, the angular momentum of light may provide an effective way to manipulate the rotational quantum state of a spiral phase plate \cite{bhattacharya2007using}, and to drive the rotational degree of other mechanical oscillators, such as levitated nanoparticles \cite{PhysRevLett.117.123604,PhysRevLett.121.033602,Ahn2020} and integrated optical waveguides \cite{Fenton:18,Hee1600485}.
Despite these exciting developments, the challenge remains in achieving strong angular momentum optomechanical interaction between light and torsional oscillators with macroscopic length scale \cite{PhysRevLett.117.123604,PhysRevLett.121.033602,Ahn2020,Fenton:18,Hee1600485}, which would lead to interesting macroscopic quantum phenomena and extend the research scope of quantum optomechanics.
In this Letter, we theoretically propose an optomechanical system with strong exchange interaction between spin angular momentum of light and a long (length $L\approx 0.1 \mathrm{mm}-1\mathrm{cm}$) torsional oscillator of optical anisotropy and demonstrate
the effectively cooling, squeezing and Schrödinger catlike state \cite{PhysRevA.55.3184} preparation of the torsional degree of freedom of
the torsional oscillator. The torsional motion of the optical anisotropic torsional oscillator induces cross optomechanical coupling between two orthogonal optical modes and the torsional motion mode.
With feasible experimental parameters, the optomechanical
coupling strength can reach the order of 20$\mathrm{kHz}$,
which can be further amplified into the strong coupling regime \cite{PhysRevLett.123.113601} (i.e. optomechanical coupling coefficient $\chi\geq 1$) by using a coherent
laser to pump one of the orthogonal polarization mode.
This enables us to control the
torsional quantum state of the torsional oscillator using only a few photons of the other optical mode, even though the original optomechanical coupling strength does
not satisfy the usual single-photon strong coupling condition \cite{PhysRevLett.107.063602}. Based on this system, it is feasible to prepare the torsional oscillator  into a
coherent superposition state of two macroscopically distinct torsional quantum states, thus provides a platform to explore the fundamental problems related with quantum torsional motion, such as the macroscopic decoherence, rotational friction and diffusion of a long quantum torsional oscillator \cite{PhysRevA.94.052109, PhysRevLett.121.040401}.
Therefore, our system greatly enriches the toolbox of quantum optomechanics.

\begin{figure}
\centering
\includegraphics[width=8.8cm]{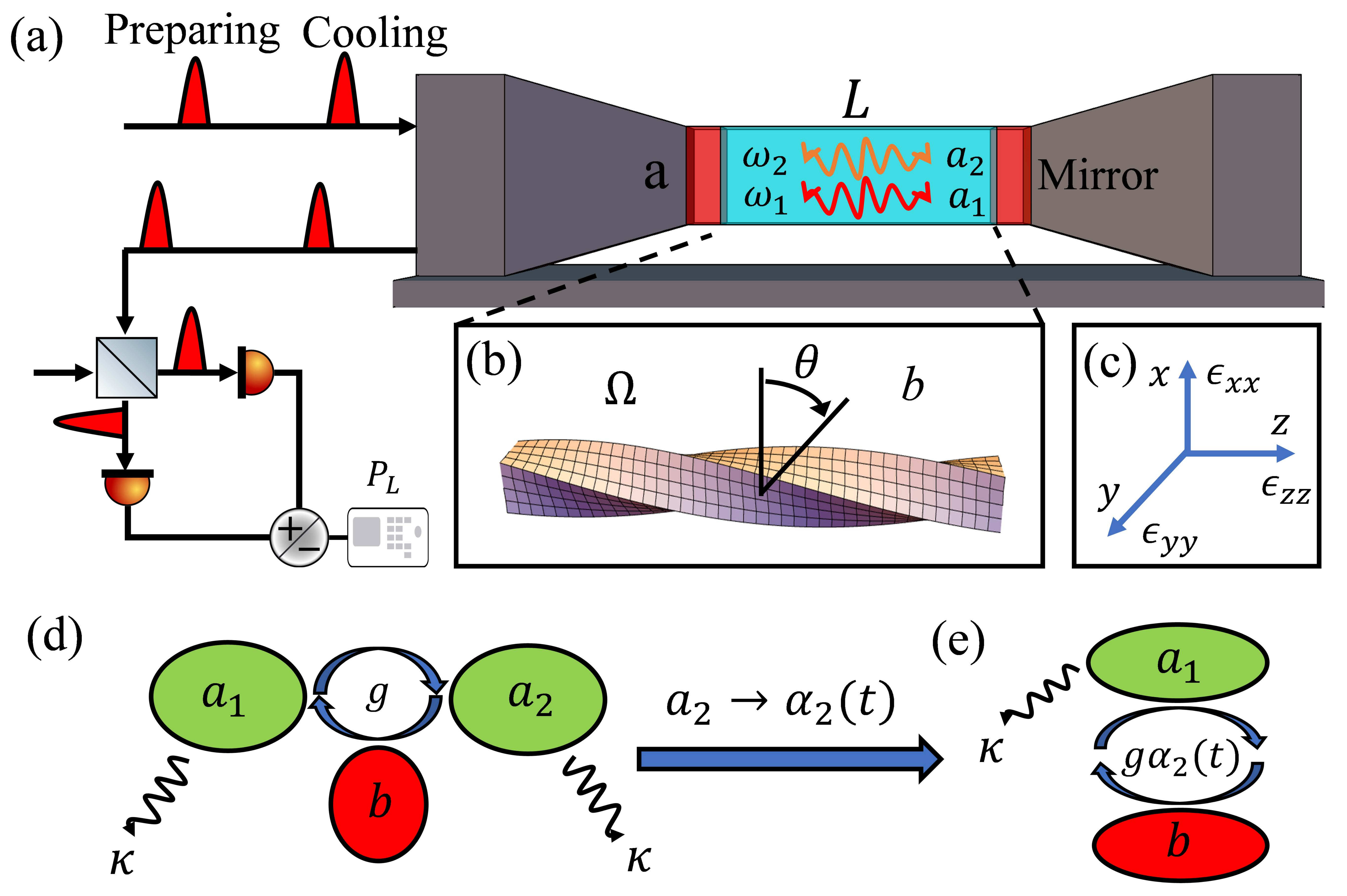}
\caption{({\bf a}) Sketch showing the experimental embodiment of this optomechanical resonator. The length of optomechanical resonator is $L$ and the width is $a$. The frequency of mechanical mode $b$ is $\Omega$ and the frequencies of the optical modes $a_{1}$ and $a_{2}$ are $\omega_{1}$ and $\omega_{2}$, respectively. ({\bf b}) The torsional mode of the optomechanical resonator with the relative permittivity $\epsilon_{ii}$  ({\bf c}) in the $i$ ($i=x,y,z$) direction. ({\bf d}) The optomechanical resonator couples with the two optical modes, the optomechanical coupling strength of which is denoted as $g$, see Eq.\eqref{HamOM}. ({\bf e}) shows the effective coupling between the torsional mode b and the optical mode $a_{1}$ when the optical mode $a_{2}$ is classically driven by a coherent laser pulse with the characteristic parameter $\alpha_2(t)$.}
\label{fig:model}
\end{figure}
{\it The model.---}The optomechanical resonator that we consider is a suspended square beam \cite{Stievater15,Keloth17,mi9110541} with length $L$ and width $a$ ($L\gg a$), the basic structure of which is presented in Fig.\ref{fig:model}. The optical anisotropy of this optomechanical resonator results in the following relative permittivity tensor $\epsilon$
\begin{equation}
\epsilon=\begin{pmatrix}
	\epsilon_{xx}& 0 & 0 \\
	0  & \epsilon_{yy} & 0  \\
	0& 0 & \epsilon_{zz}
	\end{pmatrix},
\end{equation}
where $\epsilon_{ii}$ is the relative permittivity in the $i$ ($i=x,y,z$) direction in laboratory frame. This optomechanical resonator supports two fundamental orthogonal quasi-linear polarization optical modes \cite{supp}
with the corresponding electric field profiles  $ \mathbf{E}_{1}(\mathbf{x})=\mathbf{E}_{1}(x,y)\cos(\beta_{1}z)\exp(i\omega_{1}t)$ and $ \mathbf{E}_{2}(\mathbf{x})=\mathbf{E}_{2}(x,y)\cos(\beta_{2}z)\exp(i\omega_{2}t)$ \cite{le2009cavity,le2010effect}, where $\beta_{1}$ and $\beta_{2}$ are the propagation constants of $\mathbf{E}_{1}(\mathbf{x})$ and $\mathbf{E}_{2}(\mathbf{x})$, respectively.
The total electric field operator can be written as $\mathbf{E}(\mathbf{x})=\mathbf{E}_{1}(\mathbf{x})a_{1}+\mathbf{E}_{2}(\mathbf{x})a_{2}+h.c.$, where $a_{1}$ and $a_{2}$ are the annihilation operators of two optical modes, which satisfy bosonic commutation relation $[a_{i},a_{j}^{\dagger}]=\delta_{ij}$ ($i,j=1,2$).
From a mechanical point of view, the torsional wave in a mechanical oscillator with $L\gg a$ can be described by one-dimensional Webster-type wave equation \cite{wuttke2013optically,4020}
\begin{equation}
c_{t}^{-2} \partial_{t}^{2} \phi(z,t)-\partial_{z}^{2} \phi(z,t)-\left(\frac{\partial_{z} I_{p}(z)}{I_{p}(z)}\right) \partial_{z} \phi(z,t)=0 \label{Web},
\end{equation}
where $I_{p}(z)=\int_{A} r^{2} d A$ is the polar moment of inertia at the cross section $A(z)$, $c_{t}$ is the phase velocity of torsional wave, $\phi(z,t)=\theta(z)\cos(\Omega t)$ is the amplitude of angular displacement with $\Omega$ the torsional resonance frequency and $\theta(z)$ the spatial mode function of torsional motion. In order to decrease mechanical dissipation \cite{wuttke2013optically,wuttke2014thermal}, we will focus on the fundamental mode of torsional motion which can be supposed as $\theta(z) \propto \cos(k_{t}z)$ \cite{supp,wuttke2014thermal} where $k_{t}=\Omega/c_{t}$.
By introducing phonon annihilation operator $b$ and creation operator $b^{\dagger}$, we can express the Hamiltonian of the torsional motion as $H_{\mathrm{m}}=\hbar \Omega (b^{\dagger}b+1/2)$ \cite{wuttke2014thermal}. The zero point angular displacement is given by $\theta_{zp}=\sqrt{\hbar/(2I_{\mathrm{eff}}\Omega)}$, where $I_{\mathrm{eff}}$ is the effective moment of  inertia.
The Hamiltonian of electromagnetic field can be written as $H_{\mathrm{em}}=1/2\int_{V}dV(\epsilon_{0}\mathbf{E}(\mathbf{x})\cdot \epsilon(\mathbf{x})\mathbf{E}(\mathbf{x})+\mu_{0}\mathbf{H}(\mathbf{x})\cdot \mu(\mathbf{x})\mathbf{H}(\mathbf{x}))
$ \cite{jackson1999classical}. The torsional motion of the optomechanical resonator will influence the relative permittivity $\epsilon$, and thus affect the Hamiltonian $H_{\mathrm{em}}$. If we neglect opto-elastic effect \cite{yariv1984optical,zoubi2016optomechanical} and moving boundary effect \cite{johnson2002perturbation,zoubi2016optomechanical}, the relative permittivity $\epsilon(\theta)$ of the torsional motion can be estimated as \cite{supp}
\begin{equation}
\epsilon(\theta )=R(\theta)\epsilon(0) R(-\theta)
\label{theta}
\end{equation}
where $\epsilon(0)=\epsilon$ and $R(\theta)=\left(\begin{array}{ccc}
\cos (\theta) & -\sin (\theta) & 0 \\
\sin (\theta) & \cos (\theta) & 0 \\
0 & 0 & 1
\end{array}\right)$ is a rotation matrix.
Expanding $R(\theta)$ to the first order of $\theta$, we can derive the approximated expression of $\epsilon(\theta)$ as  
\begin{equation}
\epsilon(\theta)=\epsilon(0)+ \delta\epsilon \theta(z) A
\label{epsilon}
\end{equation}
where $\delta \epsilon = \epsilon_{xx}-\epsilon_{yy}$ and $A=\begin{pmatrix}
0	& 1 & 0 \\
1	& 0 & 0 \\
0	&0  & 0
\end{pmatrix}$.
The matrix $A$ mixes two optical modes $a_{1}$ and $a_{2}$ in Hamiltonian $H_{\mathrm{em}}$ and induces exchange interaction between the optical modes $a_{1}$ and $a_{2}$. Finally, we can obtain the total optomechanical Hamiltonian as
\begin{equation}
\begin{split}
H_{\mathrm{OM}}&=\hbar \Omega b^{\dagger}b + \hbar \omega_{1} a^{\dagger}_{1}a_{1}
  +\hbar \omega_{2} a^{\dagger}_{2}a_{2}  \\& + \hbar g (b+b^{\dagger})a_{1}^{\dagger}a_{2}+\hbar g^{*} (b+b^{\dagger})a_{2}^{\dagger}a_{1}
\end{split}
\label{HamOM}
\end{equation}
where $g$ is the optomechanical coupling strength.
We note that the effective Hamiltonian of the similar form can also be found in coupled-cavity optomechanical systems \cite{PhysRevLett.109.063601,Burgwal_2020,PhysRevA.87.013839} with promising applications powered by
the single-photon non-linearities optomechanical interaction \cite{PhysRevA.87.013839} and enhanced quantum non-linearities optomechanical interaction \cite{PhysRevLett.109.063601}.

The coupling strength $g$ plays a vital role in quantum optomechanics. In the present system, $g$ can be estimated as \cite{supp}
\begin{equation}
g \propto \theta_{zp}\sqrt{\omega_{1}\omega_{2}}\left(\dfrac{\delta\epsilon}{L}\right)\int_{-L/2}^{L/2} \theta(z)\cos(\beta_{1}z)\cos(\beta_{2}z)dz.
\label{g1}
\end{equation}
The coupling strength $g$ is proportional to the optical anisotropy $\delta \epsilon$ of the torsional oscillator, see Eq.\eqref{epsilon}. The integral factor in Eq.\eqref{g1} is determined by the wave vectors $\beta_{1}$ and $\beta_{2}$ of the two optical modes and the wave vector $k_{t}$ of the torsional mode. We remark that the other optomechanical coupling mechanisms which are widely involved in integrated optomechanical systems \cite{chan2011laser,schafermeier2016quantum,PhysRevLett.126.033601}, such as moving boundary coupling effect \cite{johnson2002perturbation,zoubi2016optomechanical} and opto-elastic coupling effect \cite{yariv1984optical,zoubi2016optomechanical}, will also contribute to $g$. Nevertheless, our analysis detailed in Supplementary Material \cite{supp} shows that for the torsional motion considered in our system, the contributions of these effects are much less significant than the contribution of the optical anisotropy and can be neglected in the present system.

\iffalse
In order to understand this optomechanical interaction further, we consider a more physical and simpler example: we use a laser beam with two orthogonal linear polarization modes $a_{1}$ and $a_{2}$ to interact with an optical anisotropic membrane with width $d \ll \lambda$ and then calculate the optomechanical Hamiltonian.
Generally speaking, the spin angular momentum related optical torque $\tau_{\text{spin}} \propto M_{zz}^{\text{spin}} (z=-d/2)- M_{zz} ^{\text{spin}} (z=d/2) \propto a_{1}^{\dagger}a_{2} + h.c.$ \cite{allen2003optical}, where $ M_{zz}^{\text{spin}}$ is the optical spin angular momentum flux. The corresponding spin angular momentum optomechanical Hamiltonian has the following form
\begin{equation}
H_{\text{spin}}=-\tau_{\text{spin}} \hat{ \theta }\propto ( ga_{1}^{\dagger}a_{2} + h.c.)(b+b^{\dagger}).
\end{equation}
Essentially, our optomechanical system can be viewed as the stacking of several optical anisotropic dielectric membranes. Therefore, the optomechanical Hamiltonian of our system also has the form $H_{\text{OM}}=(g a_{1}^{\dagger}a_{2} + h.c.)(b+b^{\dagger})$, which is very different from other optomechanical Hamiltonians.

Therefore, we can conclude that the present system is feasible within the state-of-the-art experimental capability. In Tab. \ref{tab1}, we list all of the experimentally achievable parameters that we will use in numerical analysis. It can be seen that our optomechanical system works in deep unresolved sideband regime $\kappa \gg \Omega$, where $\kappa$ is the single side optical dissipation rate.
%
\fi

%
\begin{table*}
\caption{Parameters of the optomechanical system}
\begin{ruledtabular}
\begin{tabular}{ccccccccc}
$\lambda / \mathrm{nm}$ & $Q_{o}$ & $\kappa/\mathrm{MHz}$
& $\tau^{-1} / \mathrm{MHz}$
& $\Omega/ \mathrm{kHz}$ & $Q_{m}$ &$I_{\mathrm{eff}} / \mathrm{kg\cdot m^{2}}$ & $\theta_{zp}$ & $\mathrm{g/kHz}$   \\ \hline
1550 &$8.4\times 10^{5}$&$2\pi \times 225$& $ 2\pi \times 15$ &$ 2\pi \times 500 $ &$1\times 10^{4}$ & $4.4\times 10^{-25}$ & $6.16\times 10^{-9}$ & 22\\
\end{tabular}
\end{ruledtabular}
\label{tab1}
\end{table*}

{\it Single-photon optomechanical coupling.---} To achieve strong optomechanical coupling at the single-photon level, we set two optical modes $a_{1}$ and $a_{2}$ as degenerate with $\omega_{1}=\omega_{2}$ \cite{Wang:20}. Then the Hamiltonian $H_{\mathrm{OM}}$ in Eq.\eqref{HamOM} can be expressed in interaction picture as $H_{\mathrm{OM}}=\hbar g (b e^{-i \Omega t}+b^{\dagger} e^{i\Omega t})(a_{1}^{\dagger}a_{2}+a_{2}^{\dagger}a_{1})$. Using a strong coherent laser to pump the optical mode $a_{2}$, the intracavity operator $a_{2}$ can be treated classically and the Hamiltonian $H_{\mathrm{OM}}$ becomes
\begin{equation}
H_{\mathrm{OM}}^{\mathrm{new}}=\hbar g\alpha_{2}(t) (b e^{-i \Omega t}+b^{\dagger} e^{i\Omega t})(a_{1}^{\dagger}+a_{1}),
\label{eq:newH}
\end{equation}
where $\alpha_{2} = \langle a_{2} \rangle$, $\int |\alpha_{2}(t)|^{2}dt \approx 4N_{\text{in}}/\kappa$ \cite{supp}, $N_{\text{in}}$ is the total input photon number of a single coherent pulse and $\kappa$ is the decay rate of the cavity.  Without loss of generality, we assume that $\alpha_{2}$ is a real number.
By defining $\hat{x}_{L}=(a_{1}^{\dagger}+a_{1})/\sqrt{2}$, $\hat{p}_{L}=i(a_{1}^{\dagger}-a_{1})/\sqrt{2}$, $\hat{\theta}_{M}=(b^{\dagger}+b)/\sqrt{2}$, $\hat{L}_{M}=i(b^{\dagger}-b)/\sqrt{2}$, the Hamiltonian $H_{\mathrm{OM}}^{\mathrm{new}}$ becomes
$
H_{\mathrm{OM}}^{\mathrm{new}}=2\hbar g \alpha_{2}(t) \hat{x}_{L} [ \hat{\theta}_{M}\cos(\Omega t)+\hat{L}_{M}\sin(\Omega t)]$.
We consider using a short laser to pump the optical mode $a_2$, which carries $N_{\text{in}}$ photons within the duration $\tau$. If the condition $\Omega \ll \tau^{-1} \ll \kappa$ is satisfied, we can resonably neglect the second term of the Hamiltonian $H_{\mathrm{OM}}^{\mathrm{new}}$ and obtain the following effective Hamiltonian as \cite{supp}
\begin{equation}
H_{\mathrm{eff}}= 2\hbar g\alpha_{2}(t)\hat{x}_{L}\hat{\theta}_{M}.
\end{equation}
This above Hamiltonian $H_{\mathrm{eff}}$ describes the effective interaction between the optical mode $a_{1}$ and the torsional mode $b$. Because of the short pulse characteristic of $\alpha_{2}(t)$, we can define an optomechanical coupling coefficient $\chi=4g\sqrt{\int |\alpha_{2}(t)|^{2}dt}/\sqrt{\kappa}\approx 8g\sqrt{N_{\text{in}}}/\kappa$ \cite{supp}. More generally, using the optical input-output relation $a_{1}^{\mathrm{in}}(t)+a_{1}^{\mathrm{out}}(t)=\sqrt{\kappa}a_{1}(t)$, we can derive the input-output relations \cite{Bennett_2016,Vanner2013,Vanner16182,hoff2016measurement,Julsgaard2004,supp}
\begin{equation}
\begin{aligned}
\hat{x}_{L}^{\mathrm{out}}&=\hat{x}_{L}^{\mathrm{in}},\\
\hat{p}_{L}^{\mathrm{out}}&=\hat{p}_{L}^{\mathrm{in}}-\chi \theta_{M}^{\mathrm{in}},\\
\hat{\theta}_{M}^{\mathrm{out}}&=\hat{\theta}_{M}^{\mathrm{in}},\\
\hat{L}_{M}^{\mathrm{out}}&=\hat{L}_{M}^{\mathrm{in}}-\chi \hat{x}_{L}^{\mathrm{in}}.
\end{aligned}
\label{inout}
\end{equation}
Through this optomechanical interaction, the optical quadratures $\hat{x }_{L}$ and $\hat{p}_{L}$ partially exchanges information with the torsional quadratures $\hat{ \theta }_{M}$ and $\hat{L}_{M}$.Therefore, the quantum state of torsional oscillator becomes correlated with the quantum state of optical mode $a_{1}$.
It can also be seen that, the quantum state of the optical mode $a_{1}$ is unaffected by the pumping laser of mode $a_{2}$. As a result, when $\chi\geq 1$, Eq.(\ref{inout}) allows an effective coherent control of the torsional oscillator by very few photons of the optical mode $a_1$, and thus provides a new platform to realize single photon optomechanics \cite{PhysRevLett.107.063602}, even though this system do not satisfy the usually single photon strong coupling condition $g\geq\Omega$ \cite{PhysRevLett.107.063602}. The requirement for the realization of single photon optomechanics, namely to achieve $\chi\geq 1$ is $N_{\text{in}}\geq 6.45\times 10^{7}$, which is feasible in the state-of-the art experiment \cite{Vanner2013,PhysRevLett.123.113601}.

As powered by the strong single-photon optomechanical coupling with $\chi\geq1$, the torsional oscillator can be squeezed and cooled efficiently in this system using single-pulse-measurement protocol \cite{Vanner2013}. By using a single coherent pulse to pump mode $a_{2}$ and then performing a homodyne detection on quadrature $\hat{p}_{L}^{\text{out}}$, the torsional oscillator will be projected into a state that inherits the features of the input state of the optical mode $a_{1}$. As an example, we assume that the input state of the optical mode $a_{1}$ is a vacuum state, and the input state of the torsional oscillator is a thermal state with the mean phonon number $\bar{n}$. According to the input-output relation Eq.(\ref{inout}) and noticing that $\Delta\left(\hat{p}_{L}^{\text{out}}\right)$, the variance of $\hat{p}_{L}^{\text{out}}$ becomes $0$ by the homodyne detection, we obtain
\begin{equation}
\begin{aligned}
\Delta(\hat{\theta}_{M}^{\text{out}})^{2}&=\chi^{-2} \Delta\left(\hat{p}_{L}^{\text{in}}\right)^{2}+\chi^{-2} \Delta\left(\hat{p}_{L}^{\text{out}}\right)^{2}=1/(2\chi^{2}),\\
\Delta (\hat{L}_{M}^{\text{out}})^{2}&= \Delta\left(\hat{L}_{M}^{\text{in}}\right)^{2}+\chi^{2}\Delta\left(\hat{x}_{L}^{\text{in}}\right)^{2}
= (\bar{n} + 1/2)+\chi^2 /2.
\end{aligned}
\label{squeeze}
\end{equation}
Therefore, when $\chi \geq1$, $ \Delta(\hat{\theta}_{M}^{\text{out}})^{2}=1/(2\chi^{2})\leq1/2$, the $\hat{\theta}_{M}$ quadrature is thus squeezed to the level of the vacuum fluctuation and $\hat{L}_{M}$ quadrature keeps almost unaffacted. As a consequence, this torsional oscillator is squeezed into an asymmetrically cooled state (namely squeezed thermal state) \cite{Vanner2013,Vanner16182} and
has a reduced effective phonon number $\bar{n}_{\mathrm{eff}}=\left[\Delta(\hat{\theta}_{M}^{\text{out}})^{2}\Delta (\hat{L}_{M}^{\text{out}})^{2} \right]^{1 / 2}-1 / 2 \simeq\left(\bar{n} /\left(2 \chi^{2}\right)\right)^{1 / 2} \ll \bar{n}$ \cite{Vanner2013} when $\bar{n}\gg 1$ .
\begin{figure}[b]
\centering
\includegraphics[width=8cm]{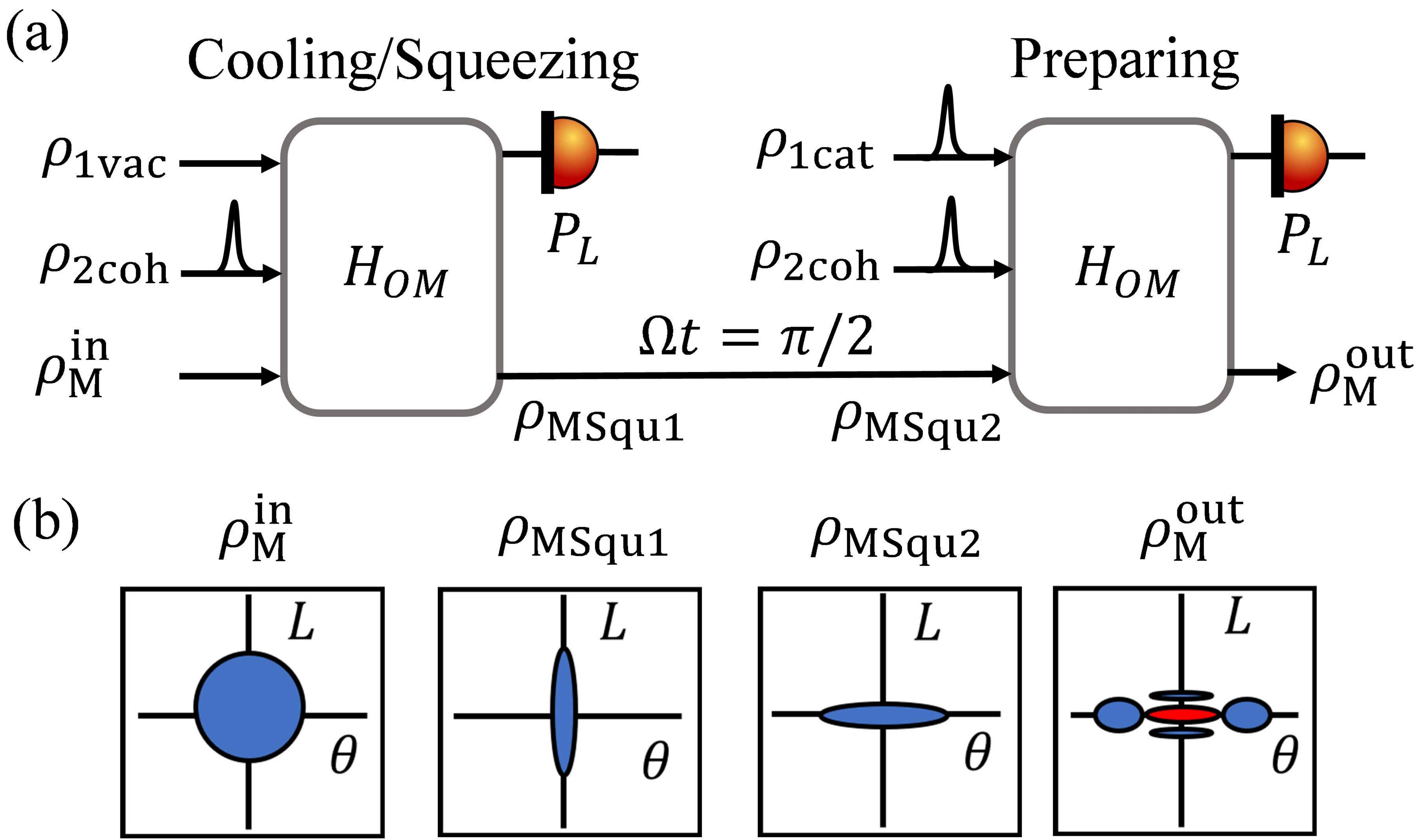}
\caption{({\bf a}) The protocol for the preparation of macroscopic Schrödinger catlike state of the torsional resonantor includes three steps: (1) the torsional resonantor is cooled and squeezed ($\rho_{\text{M}}^{\text{in}}\rightarrow \rho_{\text{MSqu1}}$) by the first optomechanical interaction $H_{OM}$ and the homodyne detection of the $\hat{p}_{L}^{\text{out}}$ quadrature of the optical mode $a_1$ ($P_L$) with the input optical state $\rho_{\text{1vac}}$ (vacuum state) and $\rho_{\text{2coh}}$ (coherent state pulse); (2) the free evolution of the torsional resonantor for time $\Omega t = \pi/2$ ($ \rho_{\text{MSqu1}}\rightarrow \rho_{\text{MSqu2}}$); (3) the torsional resonantor is prepared into a Schrödinger catlike state ($\rho_{\text{MSqu2}} \rightarrow \rho_{\text{M}}^{\text{out}}$) by the second optomechanical interaction $H_{OM}$ and the homodyne detection of the optical mode $a_1$ ($P_L$) with the input optical cat state $\rho_{\text{1cat}}$ and $\rho_{\text{2coh}}$ (coherent state pulse). ({\bf b}) illustrates the Wigner functions of the mechanical state of the torsional resonantor during the experimental protocol .}
\label{fig:exppulses}
\end{figure}
{\it Preparation of macroscopic Schrödinger catlike state.---}
We proceed to demonstrate that the torsional oscillator can be prepared in a Schrödinger catlike state using a two-pulse-measurement protocol, see Fig.\ref{fig:exppulses}. The first pulse is used to cool the torsional oscillator, as we have explained before. The second set of pulses includes a coherent laser pulse (to pump the mode $a_{2}$) and an optical catlike state pulse (i.e. the mode $a_{1}$) \cite{PhysRevA.103.013710}. The subsequent homodyne measurement on the output field $\hat{p}_{L}^{\text{out}}$ will partially transfer the quantum state from the optical mode $a_{1}$ to the torsional oscillator, thus prepare the torsional oscillator into a Schrödinger catlike state.
\begin{figure}[t]
\centering
\includegraphics[width=\linewidth]{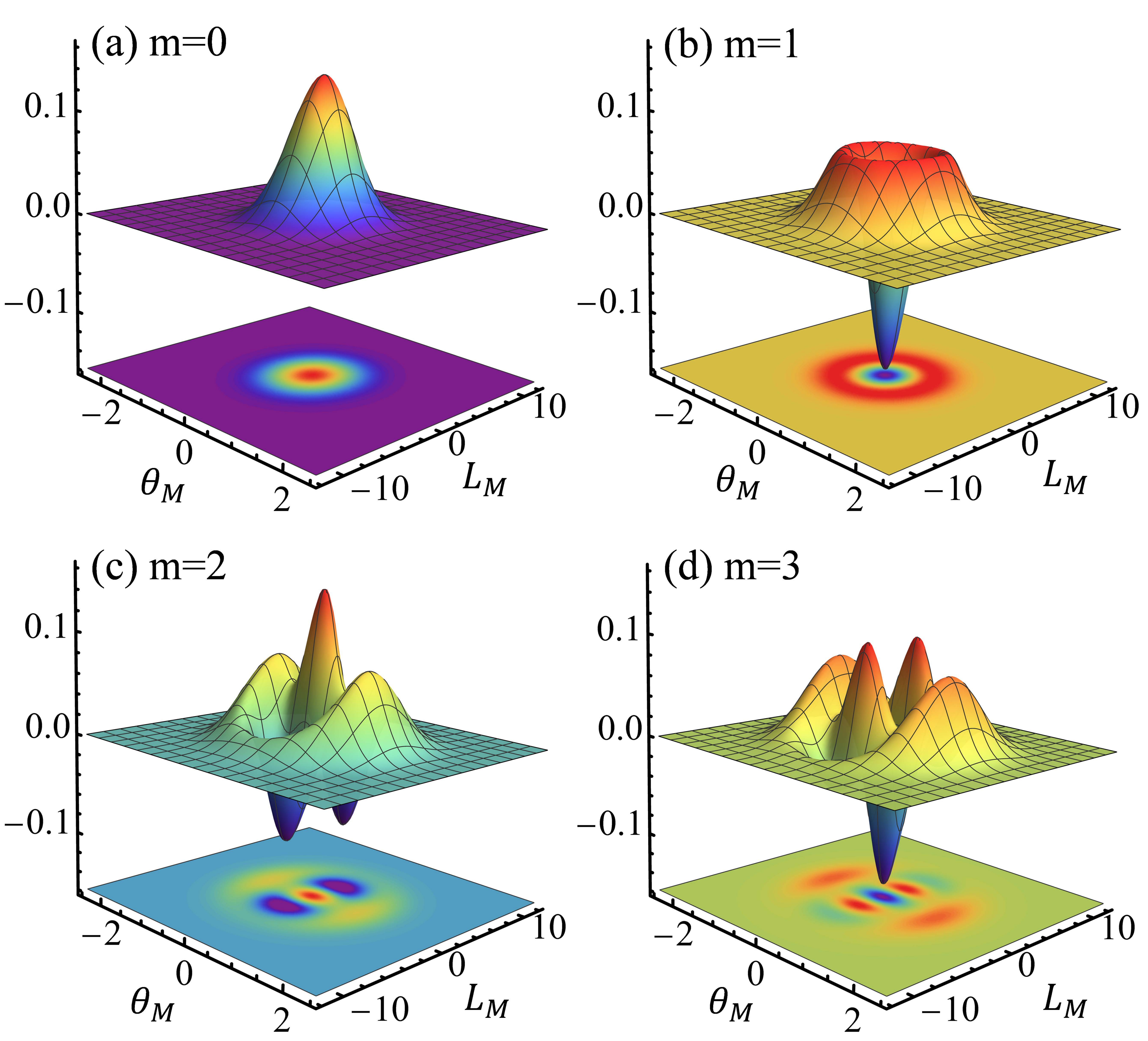}
\caption{The Wigner functions of the torsional oscillator prepared in Schrödinger catlike states with $\theta_{M}$ quadrature squeezed. In our numerical calculation, we choose the coupling coefficient $\chi=1$, the squeezing parameters $r_{1}=-r_{2}=1.15$ (i.e.10 $\mathrm{dB}$ of squeezing), $T_{\text{tap}}=(e^{2r_{1}}-1)/(e^{2r_{1}}-e^{-2r_{1}})\approx0.909$ \cite{PhysRevA.103.013710} and the environment temperature $T=100\mathrm{mK}$. The other parameters we use are listed in Table (\ref{tab1}).}
\label{fig:wigner-ps2}
\end{figure}
We use Wigner function to characterize the state of the torsional oscillator. Given the input optical cat state of mode $a_{1}$ with the Wigner function $W_{L}(x_{L},p_{L})$ and a precooled mechanical state with the Wigner function $W_{M}(\theta_{M},L_{M})$, up to a normalization factor, the Wigner function of the resulting mechanical state $W (\theta_{M} ,L_{M})$ is \cite{supp}
\begin{equation}
\begin{aligned}
W \left(\theta_{M}, L_{M}\right)=& \iint d x_{L} d p_{L} W_{L}\left(x_{L}, p_{L}+\chi \theta_{M}\right) \\
& \times W_{M} \left(\theta_{M}, L_{M}+\chi x_{L}\right) W_{\text{Homo}}(x_{L},p_{L})
\end{aligned}
\label{Wigner}
\end{equation}
where $W_{\text{Homo}}(x_{L},p_{L})=\delta(p_{L}-p)$ is the Wigner function of perfect homodyne measurement \cite{0503237} and $p$ is the outcome of the homodyne detection. Following the analysis in \cite{hoff2016measurement}, we 
choose $p=0$ in our numerical calculation. Eq.(\ref{Wigner}) describes the quantum state exchanging process between the optical mode $a_{1}$ and the torsional oscillator that is induced by the optomechanical interaction (\ref{inout}). The overlap integral between $W_{L}\left(x_{L}, p_{L}+\chi \theta_{M} \right) W_{M} \left(\theta_{M}, L_{M} +\chi x_{L}\right)$ and $W_{\text{Homo}}(x_{L},p_{L})$ describes the process of homodyne detection on the quadrature $\hat{p}_{L}^{\text{out}}$.

Comparing with the macroscopic mechanical cat states which are difficult realized in experiments, the technologies of preparing optical cat state are much more mature \cite{PhysRevA.77.062315,PhysRevLett.101.233605,Duan2019}. Here we
use the generalized photon subtraction method \cite{PhysRevA.103.013710} to prepare optical cat state. The Wigner function of the prepared optical cat state depends on two squeezing parameter $r_{1}$, $r_{2}$, the transmittance $T_{\text{tap}}$ of an asymmetric beam splitter and the subtracted photon number $m$. Using experimental feasible parameters \cite{PhysRevA.103.013710}, see Table \ref{tab1}, the Wigner functions of the resulting mechanical states for $m=0,1,2,3$ are plotted in Fig.\ref{fig:wigner-ps2}. It can be seen that when $m>0$, the Wigner function $W(\theta_{M},L_{M})$ for the output state of the torsional oscillator has negative-valued interference fringes between two significantly separated components, which demonstrates the feature of Schrödinger cat state. In addition to the mechanical Schrödinger catlike state, other non-classical quantum states, such as squeezed single phonon fock state, can also be prepared by replacing the input cat state $\rho_{1\text{cat}}$ with single photon fock state \cite{supp}.
{\it Experimental feasibility.---} The torsional oscillator can be implemented by $\alpha$-quartz \cite{2017quartz}, the optical anisotropy may reach $\delta\epsilon =0.025$ for $\lambda=1.55\mathrm{\mu m}$ \cite{GHOSH199995}. For a suspended square beam with the uniform mass density $\rho=2650\mathrm{kg/m^{3}}$, the effective moment of inertia is estimated to be $I_{\mathrm{eff}}\sim 10 I_{\mathrm{beam}}=(5/3)\rho a^{4}\int_{-L/2}^{L/2} \theta(z)^{2}dz=(5/6)\rho L a^{4}$ \cite{wuttke2014thermal}. The fundamental mode resonance frequency $\Omega$ of a torsional oscillator is in the range of $100\mathrm{kHz}-1\mathrm{MHz}$ and the mechanical quality factor $Q_{\mathrm{m}}$ is $10^{3}-10^{4}$ \cite{wuttke2013optically,he2016optomechanical,PhysRevB.95.075440,2017quartz,RevModPhys.74.991}. In our numerical analysis, we set the cross section width $a=1 \mathrm{\mu m}$ and the length $L=100 \mathrm{\mu m}$. Using the experiementally feasible parameters \cite{supp}, we estimate that the coupling constant would reach $g \sim 20 \mathrm{kHz}$.

According to Eq.(\ref{g1}), if $\theta(z)= \cos(k_{t}z)$ and $|\beta_{1}-\beta_{2}|=|\Delta \beta|=k_{t}\gg L^{-1}$, we have the coupling constant $g\propto  \delta \epsilon \theta_{zp} \propto \delta \epsilon L^{-1/2}$ \cite{supp}, where $k_{t}=\Omega/c_{t}$ and the exact acoustic velocity $c_{t}$ depends on the cutting direction of this suspended beam \cite{535483,doi:10.1063/1.1576891}.
In order to research quantum physics in longer
macroscopic scale, we can increase the
length $L$ from $0.1\mathrm{mm}$ to $1\mathrm{cm}$, e.g. with a centimeter-long optical cavity on a nanofiber \cite{Keloth17}, then the coupling constant $g$ will reduce by an order of magnitude, which can be compensated by using the optical material with a higher optical anisotropy $\delta \epsilon$ as compared with $\alpha$-quartz, such as $\mathrm{TiO_{2}}$ ($\delta \epsilon \approx 1.536$) \cite{DeVore:51} and $\mathrm{YVO_{4}}$ ($\delta \epsilon \approx 0.936$) \cite{yttrium} for $\lambda=0.63\mathrm{\mu m}$. We note that the optical quality factor $Q_{\mathrm{o}}$ on the order of $10^{5}-10^{7}$ can be achieved by using photonic crystal mirror \cite{2009photonic,Keloth17,Quan11}. Therefore, it is possible to realize strong spin angular momentum optomechanical coupling for a centimeter long torsional oscillator, and prepare the torsional oscillator in Schrödinger catlike state using a nanofiber cavity with anisotropic permittivity \cite{Snyder:83} to serve as the optomechanical resonator.

{\it Conclusion.---} We have theoretically proposed a novel optomechanical system which is based on spin angular momentum exchange interaction between light and a torsional oscillator. Our analysis shows that this system can be used to manipulate the torsional quantum state of a torsional oscillator and prepare Schrödinger catlike states with macroscopic length scale. These results pave a feasible route towards the long-standing goal of interrogating quantum mechanical phenomena at the macroscopic scale. Furthermore, the combination of the present optomechanical system with solid spin system \cite{D_Urso_2011} or atomic system \cite{Jockel2015,Leong2020} may provide a new platform of hybrid quantum optomechanics, e.g. to investigate non-demolition measurement of the phonon energy \cite{D_Urso_2011}, sympathetic cooling of a mechanical oscillator \cite{Jockel2015}, creation of a robust EPR-type of entanglement \cite{RevModPhys.77.513} between atomic ensemble and a mechanical resonator \cite{PhysRevLett.102.020501}.
{\it Acknowledgements.---} We thank Dr. Ralf Betzholz for helpful discussions. This work is supported by National Natural Science Foundation of China (Grant No.~11874024, 11690032, 12047525), the Open Project Program of Wuhan National Laboratory for Optoelectronics (2019WNLOKF002), and the Fundamental Research Funds for the Central Universities.

\bibliography{reference}

\end{document}

\title{Supplementary}  
\maketitle
\tableofcontents{\newpage}

\section{Angular Momentum Optomechanical Interaction}
In this section we discuss the optomechanical Hamiltonian. Firstly, we derive the classical Hamiltonian that represents the mutual influence of the mechanical motion and the electromagnetic field in a bounded dielectric medium, which are characterized by the electric field $\mathbf{E}(\mathbf{x})$ and the mechanical displacement field $\mathbf{Q}(\mathbf{x})$, respectively. 
For simplify we assume that $\mathbf{Q}(\mathbf{x})=\alpha \mathbf{u}(\mathbf{x})$, where $\mathbf{u}(\mathbf{x})$ is the mode function of the mechanical displacement, which satisfies the condition that $\max|\mathbf{u}|=1$ and $\alpha$ is the amplitude of the mechanical displacement. Usually $\alpha$ is much smaller than the scale of the mechanical oscillator.
By quantizing the classical electromagnetic Hamiltonian, we can derive the optomechanical Hamiltonian.

\begin{figure}[h]
	\centering
	\includegraphics[width=0.6\linewidth]{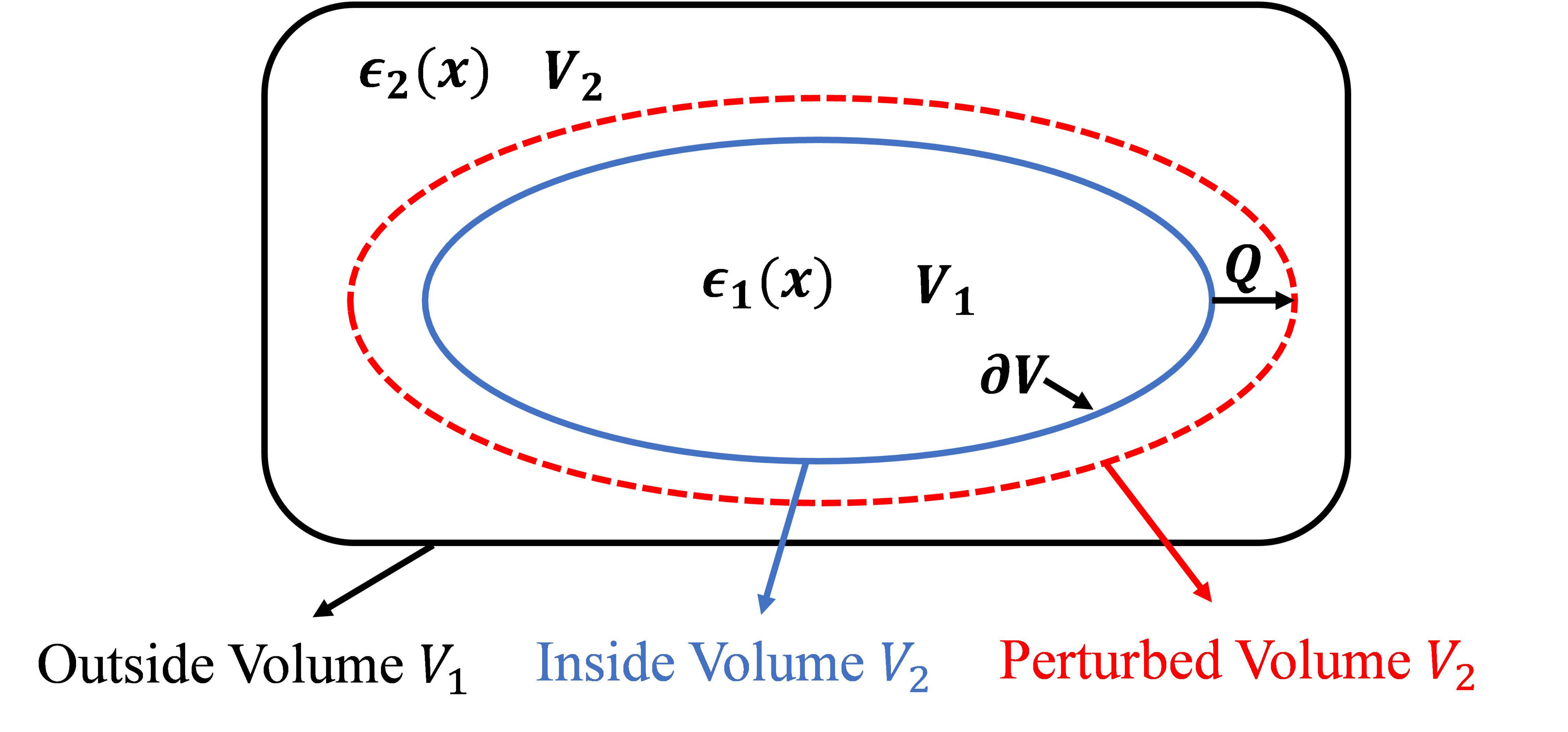}
	\caption{ A dielectric material with permittivity $ \epsilon_{1}$ in a volume $V_{1}$ (blue circle)  is  surrounded by a medium with permittivity $ \epsilon_{2}$ occupying the complementary volume $V_{2}$ (black circle). The perturbed volume is represented by the dashed red circle and $\mathbf{Q}$ is the corresponding mechanical displacement. $\partial V$ is the boundary between $V_{1}$ and $V_{2}$.}
	\label{fig:permittivity}
\end{figure}

We consider a dielectric material with permittivity $\epsilon(\mathbf{x})=\epsilon_{1}$ in a volume $V_{1}$, which is localized in a surrounding medium with permittivity $\epsilon(\mathbf{x})=\epsilon_{2}$ occupying the complementary volume $V_{2}$. For a dielectric material in vacuum $\epsilon_{2}=\epsilon_{0}$. The unperturbed permittivity can be written as
\begin{equation}
\epsilon(\mathbf{x})=\epsilon_{2}+\left(\epsilon_{1}-\epsilon_{2}\right) \Theta(\mathbf{x})
\end{equation}
where $\Theta(\mathbf{x})$ is a step function defined by
\begin{equation}
\Theta(\mathbf{x})=\left\{\begin{array}{lll}
1 & \text { for } & \mathbf{x} \in V_{1} \\
0 & \text { for } & \mathbf{x} \in V_{2}
\end{array}\right.
\end{equation}

The mechanical motion of the dielectric material will influence the total permittivity $\epsilon(\mathbf
{x})$.
We assume that $\epsilon_{1}=\epsilon_{1}(\mathbf{x},\mathbf{Q}(\mathbf{x}))=\epsilon_{1}(\mathbf{x},\alpha)$, which is also dependent on $\mathbf{Q}(\mathbf{x})$. Notice that when $\alpha=0$, $\left.\epsilon_{1}(\mathbf{x},\alpha)\right|_{\alpha=0}=\epsilon_{1}(\mathbf{x},0)$, which is the unperturbed value. 
The perturbed permittivity induced by mechanical displacement $\mathbf{Q}$ can be written as
\begin{equation}
\epsilon(\mathbf{x},\alpha)=\epsilon_{2}+\left[\epsilon_{1}(\mathbf{x},\mathbf{Q})-\epsilon_{2}\right] \Theta(\mathbf{x}+\mathbf{\mathbf{Q}})=\epsilon_{2}+\left[\epsilon_{1}(\mathbf{x},\mathbf{\alpha})-\epsilon_{2}\right] \Theta(\mathbf{x}+\alpha\mathbf{u}).
\end{equation}
Expanding $\epsilon(\mathbf{x},\alpha) $ to the first order of $\alpha$, $\epsilon(\mathbf{x},\alpha)$ is 
\begin{equation}
\begin{aligned}
\epsilon(\mathbf{x}, \alpha) & \approx \epsilon_{2}+\left[\epsilon_{1}(\mathbf{x}, 0)-\epsilon_{2}+\left.\alpha \frac{\partial \epsilon_{1}(\mathbf{x}, \alpha)}{\partial \alpha}\right|_{\alpha=0}\right]\left[\Theta(\mathbf{x})+\alpha \mathbf{u}(\mathbf{x}) \cdot \nabla \Theta(\mathbf{x})\right] \\
& \approx \epsilon_{2}+\left[\epsilon_{1}(\mathbf{x}, 0)-\epsilon_{2}\right] \Theta(\mathbf{x})+\alpha\left(\epsilon_{1}(\mathbf{x}, 0)-\epsilon_{2}\right) \mathbf{u}(\mathbf{x}) \cdot \nabla \Theta(\mathbf{x})+\alpha \left.\frac{\partial \epsilon_{1}(\mathbf{x}, \alpha)}{\partial \alpha}\right|_{\alpha=0}\Theta(\mathbf{x}) \\
&=\epsilon(\mathbf{x})+\alpha\left(\epsilon_{M B}+\epsilon_{\text{other}}\right)
\end{aligned}
\end{equation}
where  $\epsilon_{MB}=\left(\epsilon_{1}(\mathbf{x},0)-\epsilon_{2}\right) \mathbf{u}(\mathbf{x}) \cdot \nabla \Theta(\mathbf{x})$ is the moving boundary effect ($\mathbf{u} \cdot \nabla \Theta(\mathbf{x})$) induced permittivity and $\epsilon_{other}=\left.\dfrac{\partial\epsilon_{1}(\mathbf{x},\alpha)}{\partial\alpha}\right|_{\alpha=0} \Theta(\mathbf{x})$ is other effects induced permittivity. Considering the electromagnetic Hamiltonian $H_{\mathbf{em}}$ is $H_{\mathrm{em}}=1 / 2 \int_{V} d V\left( \mathbf{E}(\mathrm{x}) \cdot \epsilon(\mathrm{x}) \mathbf{E}(\mathrm{x})+ \mathbf{H}(\mathrm{x}) \cdot \mu(\mathrm{x}) \mathbf{H}(\mathrm{x})\right)$,
we can easily derive the optomechanical coupling Hamiltonian $H_{MB}$ and $H_{\text{other}}$,
\begin{equation}
	H_{MB}=\frac{1}{2} \alpha \int d V \mathbf{E}(\mathbf{x})^{\dagger} \epsilon_{MB}\mathbf{E}(\mathbf{x}),
\end{equation}
\begin{equation}
H_{\text{other}}=\frac{1}{2} \alpha \int d V \mathbf{E}(\mathbf{x})^{\dagger} \epsilon_{\text{other}}\mathbf{E}(\mathbf{x}).
\end{equation}
In following subsections we will mainly discuss $H_{MB}$ and $H_{\text{other}}$.

\subsection{Moving Boundary Coupling Effect}
If the boundary of an optical resonator is perturbed by the motion of a mechanical resonator, the energy of the total electromagnetic field will be perturbed, which can be understood from the expression $\epsilon_{MB} \propto \mathbf{u}(\mathbf{x}) \cdot \Theta(\mathbf{x}) $. By means of simple mathematical transformations, $H_{MB}$ can be expressed as 
\begin{equation}
 H_{MB}=\frac{1}{2} \alpha \int_{\partial V} d \mathbf{A} \cdot \mathbf{u}(\mathbf{x}) 
\left\{-\mathbf{E}_{\|}(\mathbf{x})^{\dagger} \Delta\epsilon \mathbf{E}_{\|}(\mathbf{x}) + \mathbf{D}_{\perp}(\mathbf{x})^{\dagger} \Delta(\epsilon^{-1}) \mathbf{D}_{\perp}(\mathbf{x})\right\}
\end{equation}
where $\Delta \epsilon=\epsilon_{1}-\epsilon_{2}$, $\Delta\left(\epsilon^{-1}\right)=\epsilon_{1}^{-1}-\epsilon_{2}^{-1}$, $\mathbf{E}_{\|}$ is the electric field which is parallel with the surface of $V_{1}$ and $\mathbf{D_{\perp}}$ is the electric displacement field which is perpendicular with the surface of $V_{1}$ . The detailed calculation process can be found in references \cite{PhysRevE.65.066611,PhysRevA.94.053827}.

\subsection{Opto-elastic Coupling Effect and Optical Material Anisotropy Coupling Effect}
In this section we mainly discuss the permittivity $\epsilon_{other}=\left.\dfrac{\partial\epsilon_{1}(\mathbf{Q})}{\partial\alpha}\right|_{\alpha=0} \Theta(\mathbf{x})$. Function $\Theta(\mathbf{x})$ shows that $\epsilon_{\text{other}}$ can only be non-zero in the region of $V_{1}$, which means that Hamiltonian $H_{\text{other}}$ is concentrated in the region of optomechanical  resonator. 

\subsubsection{Optical Material Anisotropy Coupling Effect}
We assume that the material of the optical resonator is optical anisotropy, when mechanical displacement $\alpha=0$, the unperturbed permittivity $\epsilon_{1}$ can be expressed as a diagonal matrix 
\begin{equation}
\epsilon_{1}(\mathbf{x},0)=
	\begin{pmatrix}
\epsilon_{xx}	& 0 &0  \\ 
0	& \epsilon_{yy} & 0 \\ 
0	& 0 & \epsilon_{zz}
	\end{pmatrix},
\end{equation}
where $\epsilon_{ii}$ (i=x,y,z) is the permittivity along $i$ axis.
In our main article, the displacement vector $\mathbf{u}(\mathbf{x})$ of the torsional motion is given by $\mathbf{u}(\mathbf{x})=\theta(z) (r/R) \hat{\boldsymbol{\theta}}=$
$\theta(z)(-y \hat{\boldsymbol{x}}+x \hat{\boldsymbol{y}})/R$ in cylindrical and Cartesian coordinates, respectively. Here $R$ makes $\mathbf{u}$ satisfies the condition $\max|\mathbf{u}|=1$ and $\theta(z)$ is the the spatial mode function of torsional motion and satisfies the condition $\max|\theta(z)|=1$.

In lab frame, the rotation of the optical material will influence $\epsilon_{1}$,
\begin{equation}
	\epsilon_{1}(\mathbf{x},\alpha)=R(\alpha \theta(z)) \epsilon_{1}(0) R(-\alpha \theta(z)),
	\label{rotation per}
\end{equation}
where $R(\alpha)$ is a rotation matrix,
\begin{equation}
R(\alpha)=\left(\begin{array}{ccc}
\cos (\alpha) & -\sin (\alpha) & \\
\sin (\alpha) & \cos (\alpha) & \\
0 & 0 & 1
\end{array}\right).
\end{equation}
If $\alpha \ll 1$, $R(\alpha)$ can be expanded into $R(\alpha)=I+\alpha T$, where $T$ is 
\begin{equation}
T=
	\begin{pmatrix}
	0	&-1  &0  \\ 
	1	&  0& 0\\ 
	0	&0  & 0
	\end{pmatrix}.
\end{equation}
Expanding $\epsilon_{1}(\mathbf{x},\alpha)$ to the first order of $\alpha$, we have 
\begin{equation}
	\epsilon_{1}(\mathbf{x},\alpha) \approx \epsilon_{1}(\mathbf{x},0) + \alpha\theta(z) [T,\epsilon_{1}(\mathbf{x},0)]=\epsilon_{1}(\mathbf{x},0) + \alpha\theta(z) \delta \epsilon A
\end{equation}
where $\delta \epsilon=\epsilon_{x x}-\epsilon_{y y}$ and $A=\left(\begin{array}{lll}0 & 1 & 0 \\ 1 & 0 & 0 \\ 0 & 0 & 0\end{array}\right)$.

\subsubsection{Opto-elastic Coupling Effect}
The refractive index of most material will be influenced by opto-elastic effect when material is subject to strain. Mathematically, the opto-elastic effect is described by relating the change of the permittivity tensor $\delta \eta,$ to the material strain by Pockel's tensor $p$ \cite{PhysRevB.3.2778}, 
$
(\boldsymbol{\delta} \boldsymbol{\eta})_{i j}=\boldsymbol{p}_{i j k l} s_{k l}
$,
where $s_{mn}$ is the strain tensor of the material with $s_{m n}=\frac{1}{2}\left(\frac{\partial \boldsymbol{u}_{m}}{\partial \boldsymbol{x}_{n}}+\frac{\partial \boldsymbol{u}_{n}}{\partial \boldsymbol{x}_{m}}\right)$.
The modified permittivity tensor follows from $\epsilon^{OE}=\left(\epsilon^{-1}+\alpha\delta \eta\right)^{-1}$, where $\epsilon$ is the unmodified permittivity tensor.
Considering equation (\ref{rotation per}), the final expression of $\epsilon_{1}^{OE}(\mathbf{x},\alpha)$ is 
\begin{equation}
\begin{split}
\epsilon_{1}^{OE}(\mathbf{x},\alpha)&=( \epsilon_{1}(\mathbf{x},\alpha)^{-1} + \alpha\delta \eta)^{-1}\\
&=\left( R(\alpha \theta) \epsilon_{1}^{-1}(\mathbf{x},0) R(-\alpha \theta) + \alpha \delta \eta\right)^{-1} \\
&\approx\left( (I+\alpha \theta T) \epsilon_{1}^{-1}(\mathbf{x},0) (I-\alpha \theta T) + \alpha \delta \eta\right)^{-1} \\
& \approx \left( \epsilon_{1}^{-1}(\mathbf{x},0)+  \alpha \theta (T \epsilon_{1}^{-1}(\mathbf{x},0) - \epsilon_{1}^{-1}(\mathbf{x},0) T) +\alpha \delta \eta  \right)^{-1} \\
& \approx \epsilon_{1}(\mathbf{x},0) +\alpha \theta \delta\epsilon A-\alpha \epsilon_{1}(\mathbf{x},0) \delta \eta \epsilon_{1}(\mathbf{x},0).
\end{split}
\end{equation}

\subsubsection{Expression of $\epsilon_{\text{other}}$}
The final expression of $\epsilon_{other}$ is
\begin{equation}
\begin{split}
\epsilon_{\text{other}}&=\dfrac{\partial\epsilon_{1}(\mathbf{Q})}{\partial\alpha} \Theta(\mathbf{x})\\
&= \theta [T,\epsilon_{1}(\mathbf{x},0)]\Theta(\mathbf{x})- \epsilon_{1}(\mathbf{x},0) \delta \eta \epsilon_{1}(\mathbf{x},0)\Theta(\mathbf{x})\\
&=\epsilon_{MA}+\epsilon_{OE}.
\end{split}
\end{equation}
The first term $\epsilon_{MA}=\theta [T,\epsilon_{1}(\mathbf{x},0)]\Theta(\mathbf{x})=\theta \delta \epsilon A\Theta(\mathbf{x}) $ is related with the optical anisotropy of the material and the second term $\epsilon_{OE}=- \epsilon_{1}(\mathbf{x},0) \delta \eta \epsilon_{1}(\mathbf{x},0)\Theta(\mathbf{x})$ is related with opto-elastic effect. Therefore, the optomechanical Hamiltonian $H_{\text{other}}$ can be divided into two terms $H_{AM}$ and $H_{OE}$,
\begin{equation}
H_{OE}=-\frac{1}{2} \alpha \int_{V_{1}} d V \mathbf{E}(\mathbf{x})^{\dagger}  \epsilon_{1}(\mathbf{x},0) \delta \eta \epsilon_{1}(\mathbf{x},0)\mathbf{E}(\mathbf{x}),
\end{equation}
\begin{equation}
 H_{MA}=\frac{1}{2} \alpha \delta\epsilon \int_{V_{1}} d V \theta(z) \mathbf{E}(\mathbf{x})^{\dagger}  A \mathbf{E}(\mathbf{x}).
\end{equation}

\subsection{Quantization of The Optomechanical Hamiltonian }
The quantized optomechanical Hamiltonian is obtained from
the classical one by replacing the classical amplitude $\alpha$ and the
electromagnetic fields by operators. 
We define $x_{zp}=\sqrt{\hbar/(2m_{\mathrm{eff}}\Omega)}$ is the zero point linear displacement of the torsional oscillator, where $m_{\mathrm{eff}}=\int_{V}|\mathbf{u}(\mathbf{x})|^{2}\rho dV$ is the effective mass, and $\theta_{zp}=\sqrt{\hbar/(2I_{\mathrm{eff}}\Omega)}$ is the effective zero point angular displacement of the torsional oscillator and $I_{\mathrm{eff}}$ is the effective moment of inertia of the torsional oscillator.  
For a two mode electric field $\mathbf{E}=\mathbf
E_{1} a_{1}+\mathbf{E}_{2} a_{2}+h.c.$ ($\vec{E}_{1}$ and $\vec{E}_{2}$ are not normalized), the original optomechanical Hamiltonian $H_{\text{int}}$ is 
\begin{equation}
H_{\text{int}}=H_{MB}+H_{OE}+H_{MA}=\hbar(b+b^{\dagger})\sum_{i,j=1,2}a_{i}^{\dagger}a_{j} g_{ij} 
\end{equation}
where $g_{ij}=g_{ijMB}+g_{ijOE}+g_{ijAM}$ and 
\begin{equation}
\begin{aligned}
g_{ijMB}&=\frac{x_{zp}}{2} \int_{\partial V} d \mathbf{A} \cdot \mathbf{u}(\mathbf{x}) 
\left\{-\mathbf{E}_{i\|}(\mathbf{x})^{\dagger} \Delta\epsilon \mathbf{E}_{j\|}(\mathbf{x}) + \mathbf{D}_{i\perp}(\mathbf{x})^{\dagger} \Delta(\epsilon^{-1}) \mathbf{D}_{j\perp}(\mathbf{x})\right\},
\\g_{ijOE}&=-\frac{x_{zp}}{2}  \int_{V_{1}} d V \mathbf{E}_{i}(\mathbf{x})^{\dagger}  \epsilon_{1}(\mathbf{x},0) \delta \eta \epsilon_{1}(\mathbf{x},0)\mathbf{E}_{j}(\mathbf{x}),\\
g_{ijMA}&=\dfrac{1}{2}\theta_{zp}\delta\epsilon\int_{V_{1}}dV\theta(z)\left(E_{i y}(\mathbf{x}) E_{j x}^{*}(\mathbf{x})+E_{j y}^{*}(\mathbf{x}) E_{i x}(\mathbf{x})\right).
\label{gOE}
\end{aligned}
\end{equation}
Here we use "MB" to represent moving-boundary coupling effect, use "OE" to represent opto-elastic coupling effect and use "MA" to represent material anisotropy coupling effect.

\subsection{Discussion About the Torsional Oscillator}
The exact values of mode function $\theta(z)$ and resonance frequency $\Omega$ depends on the detailed geometrical structure of the torsional resonator, here we use a method to estimate the rough profile of $\theta(z)$. 
Considering a suspended slender beam with two ends fixed, the corresponding wave equation for a cylinder with varying cross-section is known as Webster-type equation and takes the form \cite{PhysRevA.88.061801}
\begin{equation}
c_{t}^{-2} \partial_{t}^{2} \phi(t, z)-\partial_{z}^{2} \phi(t, z)-\left(\frac{\partial_{z} I_{p}(z)}{I_{p}(z)}\right) \partial_{z} \phi(t, z)=0
\label{Web}
\end{equation}
where $I_{p}(z)=\int_{A} r^{2} d A$ is the polar moment of inertia at the cross section $A(z)$ and $c_{t}$ is the phase velocity of the torsional wave.
To solve the wave equation, it is convenient to first separate it in time and space and construct solutions of the form $\phi(t, z)=\theta(z) \cos (\Omega t)$, so that only the differential equation in the axial coordinate remains.

In the simplest case 
where the torsional oscillator is a homogeneous cylinder, we have $\partial_{z} I_{p}(z)=0$. Then the wave equation (\ref{Web}) will be reduced to its simplest form
\begin{equation}
\partial_{z}^{2} \theta( z) +k_{t}^{2} \theta( z)=0,
\end{equation}
with the well-known solutions:
\begin{equation}
\theta_{\mathrm{s}}(z)=\sin \left(k_{t} z\right), \quad \theta_{\mathrm{c}}(z)=\cos \left(k_{t} z\right), \quad \text { where } k_{t}=\Omega / c_{t}.
\end{equation}
At the ends of the torsional oscillator, $\theta(z)=0$, and we can derive the
resonance frequency $\Omega$ from the corresponding boundary condition. 
However, the torsional oscillator also includes other structures besides the square suspended beam and the mode function $\theta(z)$ will be more complex than sine function or cosine function, and the effective moment of inertia $I_{\mathrm{eff}}$ will also larger than $I_{\mathrm{beam}}=
1 / 6 M a^{2} \int_{-L/2}^{L/2} \theta(z)^{2} d z= 1 / 12 \rho L a^{4}$.  As a simple example, we consider 
the fundamental mode of a suspended beam. The eigen wave vector $k_{t}$ of the sine mode $\sin(k_{t}z)$ is $2\pi n/L $ with boundary condition $\theta_{s}(L/2)=0$, where $n$ is an integer, and the wave vector $k_{t}$ of the cosine function mode $\cos(k_{t}z)$ is $2\pi (n+1/2)/L $ with boundary condition $\theta_{c}(L/2)=0$.
Therefore, the fundamental mode function $\theta(z)$ of the torsional oscillator should be an even function and the piece-wise mode function of the square beam should be a cosine function $\cos(k_{t}z)$.
In our calculation, we assume $\rho=2650\mathrm{kg/m^{3}}$, $c_{t}=5000\mathrm{m/s}$ \cite{RevModPhys.74.991}, $L=100 \mathrm{\mu m}$, $a=1\mathrm{\mu m}$, $\Omega=2\pi\times 500\mathrm{kHz}$, $I_{eff}=10\times 1 / 6 \rho a^{4} \int_{-L/2}^{L/2} \theta(z)^{2} d z= 10 / 12 \rho L a^{4}=4.4163\times 10^{-25} \mathrm{ kg\cdot m^{2}}$ and $k_{t}=\Omega/c_{t}=100\pi/\mathrm{m}$. The parameters we will use is listed in Table (\ref{para}). 
\begin{table}[h]
\caption{Parameters used in calculation}	
	\begin{tabular}{|c|c|c|c|c|c|c|c|c|c|c|c|}
		\hline\hline
     		$a/\mathrm{\mu m }$ & $L/\mathrm{\mu m }$ & $\rho/\mathrm{kg\cdot m^{-3} }$ & $\Omega/\mathrm{kHz}$&$c_{t}/\mathrm{km\cdot s^{-1}}$&$I_{\text{eff}}/\mathrm{kg\cdot m^{2}}$ &$k_{t}/\mathrm{m^{-1}}$ & $\beta_{1}/k$&  $\beta_{2}/k$&$\epsilon_{xx} $ & $ \epsilon_{yy}$ & $\epsilon_{zz} $                      \\ \hline
		1&    100  &  2650&$2\pi \times 500$ & 5.0&$4.4163\times 10^{-25}$  &100$\pi$  & 1.2859  &  1.2926 &1.5326&1.5277 &1.5277 \\ \hline\hline
	\end{tabular}
	\label{para}
\end{table}

\subsection{Calculation of the Coupling Constants}
\begin{figure}[ht]
	\centering
	\includegraphics[width=0.8\linewidth]{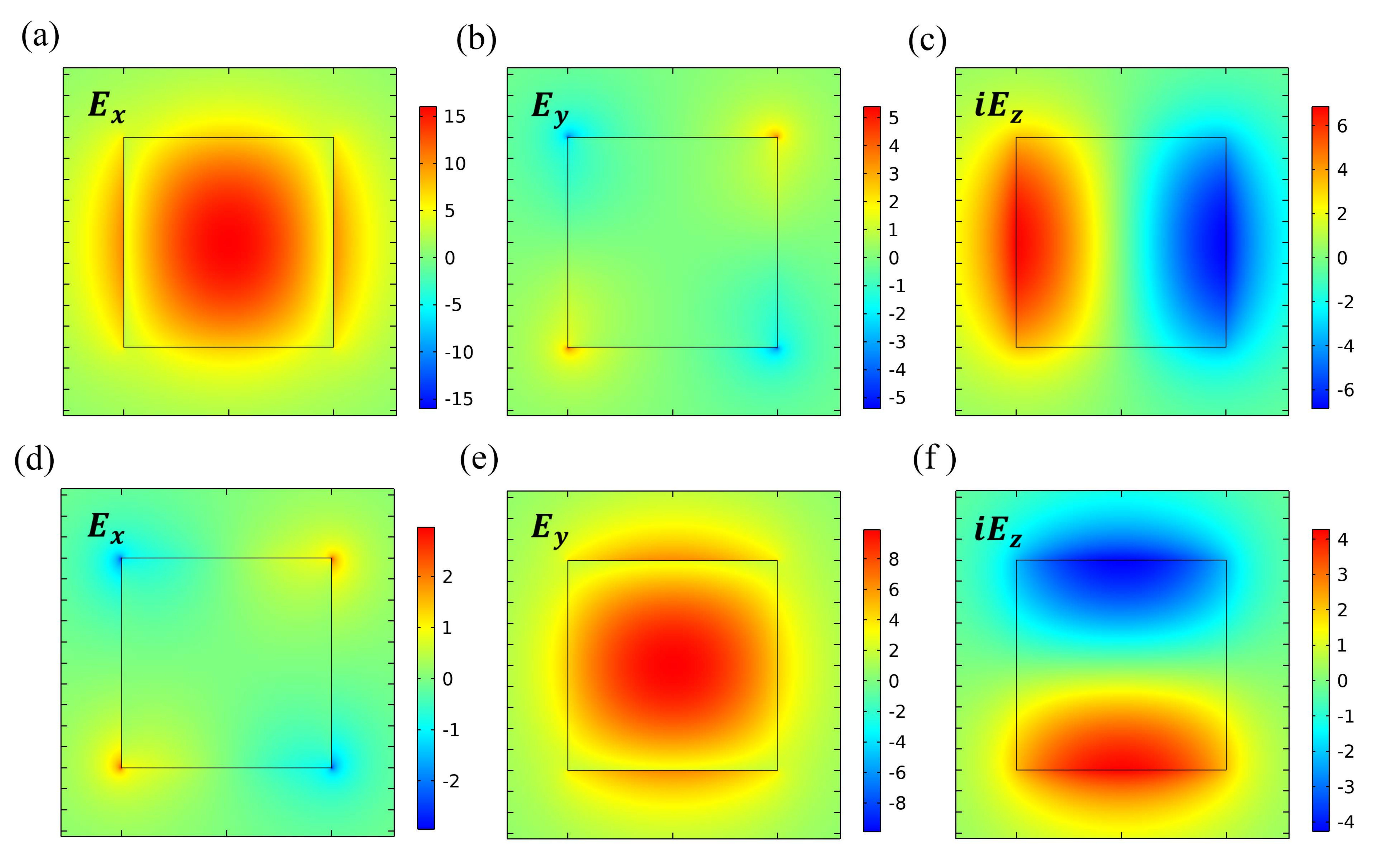}
	\caption{(a), (b) and (c) are the electric fields distribution of the TE-like mode. (d), (e) and (f) are the electric field distribution of the TM-like mode. It is obvious that $|\mathbf{E}_{i}|^{2}$ (i=x,y,z) is symmetric about $x$ axis and $y$ axis. }
	\label{fig:em-field}
\end{figure}

In this section we discuss the exact values of the coupling constants. When wavelength $\lambda=1550$nm, the relative permittivity tensor of $\alpha$-quartz are \cite{GHOS199995}
\begin{equation}
\epsilon(0)=\begin{pmatrix}
1.5326^{2}& 0 &0  \\ 
0& 1.5277^{2} &  0\\ 
0&0  & 1.5277^{2}
\end{pmatrix} ,
\end{equation}
and the corresponding mode profiles of the electric field $\mathbf{E}_{1}$ and the electric field $\mathbf{E}_{2}$ are plotted in Fig. (\ref{fig:em-field}).

\paragraph{$g_{11MB}\approx 0$ and $g_{22MB}\approx 0$} Because of the symmetry of the optical resonator, $|\mathbf{E}_{i\|}(\mathbf{x})|^{2}$ and  $|\mathbf{D}_{i\perp}(\mathbf{x})|^{2}$ (i=x,y,z) are even functions about $x$ axis and $y$ axis simultaneously. However, torsional mode function $\mathbf{u}(\mathbf{x})$ is a odd function about $x$ axis and $y$ axis. Therefore, the integration $g_{iMB}$ will approach $0$. 

\paragraph{$g_{12MB}\approx0 $} Form Fig. (\ref{fig:em-field}) we know that $\mathbf{E}_{1}(\mathbf{x})$ is a quasi-$x$ polarized electric field with $|E_{1y}| \ll |E_{1x}|$ and $\mathbf{E}_{2}(\mathbf{x})$ is a quasi-$y$ polarized electric field with $|E_{2x}| \ll |E_{2y}|$. In addition, $\mathbf{E}_{1\|}(\mathbf{x})^{\dagger} \Delta\epsilon \mathbf{E}_{2\|}(\mathbf{x})$ usually can be expanded into the summation of  $E_{1m}(\mathbf{x})^{\dagger} \Delta\epsilon E_{2m}(\mathbf{x})$ ($m=x,y,z$). Considering the normalized values of $\mathbf{E}(\mathbf{x})$ and $\mathbf{E}(\mathbf{x})$, 
\begin{equation}
g_{12MB} \propto \sum_{m=x,y,z} \frac{\int\left(- E_{1 m}^{*}\Delta \epsilon E_{2 m}+  D_{1 m}^{*}\Delta (\epsilon^{-1})  D_{2 m}\right)dA}{\sqrt{\int \mathbf{E}_{1} \cdot \epsilon \mathbf{E}_{1} d x d y} \cdot \sqrt{\int \mathbf{E}_{2} \cdot \epsilon_{2} \mathbf{E}_{2} d x d y}},
\end{equation}
Thus the value of $g_{MB}$ will approach $0$.

\paragraph{$g_{ijOE} \approx 0$} The exact values of $g_{ijOE}$ depends on the strain tensor $s_{mn}$, electric field $\mathbf{E}_{1}(\mathbf{x})$ and electric field $\mathbf{E}_{2}(\mathbf{x})$. When the mode function $\theta(z)$ of the torsional oscillator is a cosine function $\theta(z)=\cos(k_{t}z)$, the strain tensor $s_{mn}$ is 
\begin{equation}
	\mathbf{s}=\dfrac{1}{2}\sin(k_{t}z)\dfrac{k_{t}}{R}
	\begin{pmatrix}
0	&  0& y \\ 
0	& 0 &   -x  \\
y 	& - x   & 0
	\end{pmatrix} .
	\label{strain}
\end{equation}
In our main article, we assume that $\mathbf{E}_{i}(\mathbf{x})=\mathbf{E}_{i}(x,y)\cos(\beta_{i}z)$. Substituting Equ. (\ref{strain}) into Equ. (\ref{gOE}), we can easily derive the conclusion that 
\begin{equation}
	g_{ijOE} \propto \int_{-L/2}^{L/2} \sin(k_{t}z)\cos(\beta_{i}z)\cos(\beta_{j}z) dz=0.
\end{equation}
We need to note that we do not need the exact value of the Pockel's tensor p.

\paragraph{$g_{11MA} \approx 0$ and $g_{22MA} \approx 0$} The reason that $g_{iiAM} \approx 0$ is similar with the reason that $g_{12MB}=0$. 

\paragraph{$g_{12MA} \neq 0$} Considering the normalized condition of $\mathbf{E}_{1}(\mathbf{x})$ and $\mathbf{E}_{2}(\mathbf{x})$, we can express $g_{12MA}$ as
\begin{equation}
\begin{aligned}
	g_{12MA} &=\dfrac{\theta_{zp}}{2}\int_{V_{1}}dxdy\left(E_{1 y}(x,y) E_{2 x}^{*}(x,y)+E_{2 y}^{*}(x,y) E_{1 x}(x,y)\right) \times \int_{-L/2}^{L/2} \theta(z)\cos(\beta_{i}z)\cos(\beta_{i}z) dz \\
	 &=\frac{\theta_{zp}}{2}\left(\epsilon_{xx}-\epsilon_{yy}\right) \sqrt{\omega_{1} \omega_{2}}
	 \frac{\int\left(E_{2 x} E_{1 y}^{*}+E_{2 y} E_{1 x}^{*}\right) d x d y}{\sqrt{\int \mathbf{E}_{1} \cdot \epsilon \mathbf{E}_{1} d x d y} \cdot \sqrt{\int \mathbf{E}_{2} \cdot \epsilon_{2} \mathbf{E}_{2} d x d y}}
	 \times \frac{\int_{-L/2}^{L/2} \cos(k_{t}z) \cos \left(\beta_{1} z\right) \cos \left(\beta_{2} z\right) d z}{\sqrt{\int_{-L/2}^{L/2} \cos \left(\beta_{1} z\right)^{2} d z} \cdot \sqrt{\int_{-L/2}^{L/2} \cos \left(\beta_{2} z\right)^{2} d z}} \\
	 & \propto \theta_{z p} \sqrt{\omega_{1} \omega_{2}}\left(\frac{\delta \epsilon}{L}\right) \int_{-L / 2}^{L / 2} \theta(z) \cos \left(\beta_{1} z\right) \cos \left(\beta_{2} z\right) d z.
\end{aligned}
\label{g12AM}
\end{equation}
In the last step of Equ. (\ref{g12AM}), we use the approximation that $\int_{-L/2}^{L/2}\cos(\beta_{1} z)^{2}dz \approx \int_{-L/2}^{L/2}\cos(\beta_{2} z)^{2}dz \approx L/2$.

\paragraph{Discussion about $g_{12AM}$}
If $\theta(z) \propto \cos(k_{t}z)$, then the expression of $g_{12MA}$ will be
\begin{equation}
\begin{split}
     g_{12AM} & \propto \dfrac{1}{L}\int_{-L / 2}^{L / 2} \cos(k_{t}z) \cos \left(\beta_{1} z\right) \cos \left(\beta_{2} z\right) d z  \\&
     =   -\frac{\sin \left(\frac{1}{2} L\left(-\beta_{1}-\beta_{2}+k_{t}\right)\right)}{2L(\beta_{1}+\beta_{2}-k_{t})}+\frac{\sin \left(\frac{1}{2} L\left(\beta_{1}-\beta_{2}+k_{t}\right)\right)}{2L(\beta_{1}-\beta_{2}+k_{t})}+\frac{\sin \left(\frac{1}{2} L\left(-\beta_{1}+\beta_{2}+k_{t}\right)\right)}{2L(-\beta_{1}+\beta_{2}+k_{t})}+\frac{\sin \left(\frac{1}{2} L\left(\beta_{1}+\beta_{2}+k_{t}\right)\right)}{2L(\beta_{1}+\beta_{2}+k_{t})}.
\end{split}
\end{equation}
In this system, $\beta_{1}\sim 2\pi/\lambda$ and $\beta_{2}\sim 2\pi/\lambda$ where $\lambda$ is the wavelength of the electromagnetic field. If $L \geq 0.1 \mathrm{mm}$, $\beta_{1}L \gg 1$ and $\beta_{2}L \gg 1$. When $\beta_{1}-\beta_{2}=k_{t} \gg L^{-1}$, $\frac{\sin \left(\frac{1}{2} L\left(\beta_{1}-\beta_{2}+k_{t}\right)\right)}{(\beta_{1}-\beta_{2}+k_{t})L}=\dfrac{\sin(k_{t}L)}{2k_{t}L} \ll 1$ and $\frac{\sin \left(\frac{1}{2} L\left(\beta_{1}-\beta_{2}-k_{t}\right)\right)}{2(\beta_{1}-\beta_{2}-k_{t})L}=\dfrac{1}{4}$. Notice that $\theta_{zp}=\sqrt{\hbar/(2I_{\text{eff}}\Omega)}\propto L^{-1/2}$, we have $g\approx g_{12MA} \propto \delta \epsilon \theta_{z p} \propto \delta \epsilon L^{-1 / 2}$.

Using the parameters listed in Table.\ref{para}, the exact values of g is listed in the following table:
\begin{table}[h]
	\caption{Table about $g$}
	\begin{tabular}{|c|c|c|c|c|c|c|}
		\hline\hline
		$g_{12AM}/\mathrm{kHz }$ & $g_{11AM}/\mathrm{kHz }$ & $g_{22AM}/\mathrm{kHz }$ & $g_{12MB}/\mathrm{kHz }$ & $g_{11MB}/\mathrm{kHz }$ & $g_{22MB}/\mathrm{kHz }$ & $g_{ijOE}/\mathrm{kHz }$                      \\ \hline
		22&  0    &  0 & 0.081  &  -0.01 & -0.01  &0 \\ \hline\hline
	\end{tabular}
\label{g}
\end{table}

\subsection{A Simple Explanation For Our Optomechanical Hamiltonian}
Considering the energy of the torsional oscillator and neglecting the optomechanical interaction terms $a_{1}^{\dagger}a_{1}(b+b^{\dagger})$ and $a_{2}^{\dagger}a_{2}(b+b^{\dagger})$, the Hamiltonian of this system can be expressed as 
\begin{equation}
H_{\mathrm{OM}} =\hbar \Omega b^{\dagger} b+\hbar \omega_{1} a_{1}^{\dagger} a_{1}+\hbar \omega_{2} a_{2}^{\dagger} a_{2} 
+\hbar g\left(b+b^{\dagger}\right) a_{1}^{\dagger} a_{2}+\hbar g^{*}\left(b+b^{\dagger}\right) a_{2}^{\dagger} a_{1}
\label{HOM}
\end{equation}
where $g=g_{12MA}+g_{12OE}+g_{12MB} \approx g_{12AM}$.

\begin{figure}
	\centering
	\includegraphics[width=0.75\linewidth]{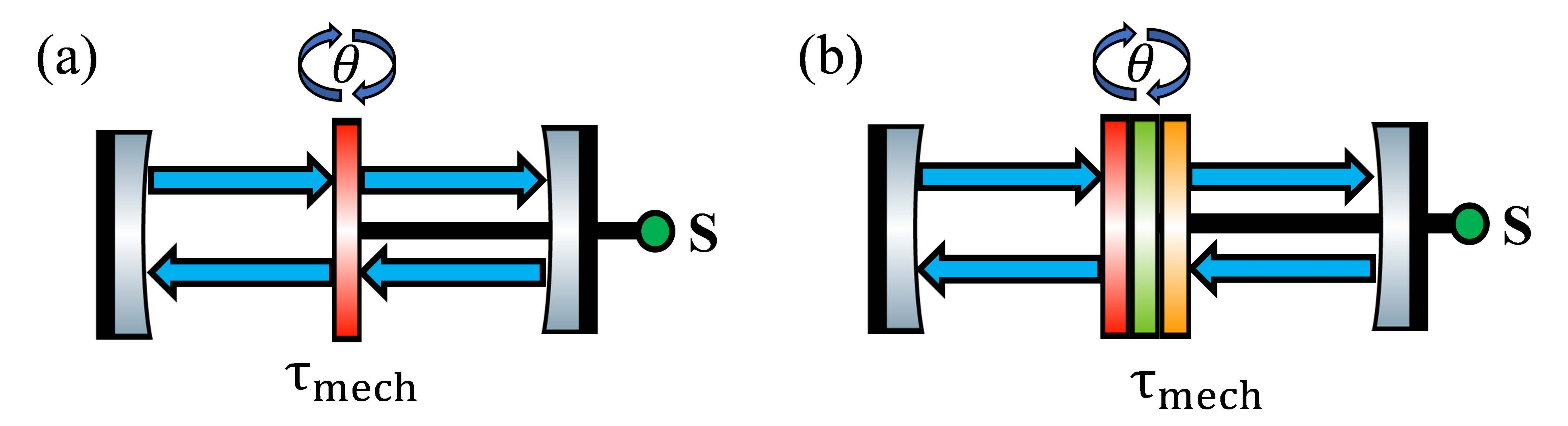}
	\caption{A simple example for our optomechanical model. (a) An extremely thin anisotropic membrane interacts with the cavity photons. Under the influencing the cavity photons, the membrane will experience a mechanical torque $\tau_{\text{mech}}$. (b) We stack several optical anisotropic dielectric membranes in this cavity. The total mechanical torque $\tau_{\text{mech}}$ will be similar with the mechanical torque $\tau_{\text{mech}}$ in (a).}
	\label{fig:example}
\end{figure}

In order to understand this optomechanical interaction further, we consider a more physical and simpler example (see Fig. \ref{fig:example}(a)): we use a laser beam with two orthogonal linear polarization modes $a_{x}$ and $a_{y}$ to interact with an optical anisotropic membrane with width $d \ll \lambda$ and then calculate the optomechanical Hamiltonian. According to the angular momentum conservation law \cite{novotny2012principles}, 
along the direction of the optical axis $z$, the mechanical torque $\mathbf{\tau}_{\mathrm{mech}}$ experienced by this membrane is 
\begin{equation}
\mathbf{\tau}_{\mathrm{mech}}=
-\mathbf{\tau}_{\mathrm{field}}
-(\int_{\partial V} \stackrel{\leftrightarrow}{\mathbf{M}}(\mathbf{r}, t)  \cdot \mathbf{n}\mathrm{d}S)_{z},
\end{equation}
where $\mathbf{\tau}_{\mathrm{field}}=\left(\frac{\mathrm{d}}{\mathrm{d} t} \frac{1}{c^{2}} \int_{V}\mathbf{r}\times(\mathbf{E} \times \mathbf{H}) \mathrm{d} V\right)_{z}$ is the electromagnetic torque, $\stackrel{\leftrightarrow}{\mathbf{M}}$ is optical angular momentum flux density and the integral $\int_{\partial V} \stackrel{\leftrightarrow}{\mathbf{M}}(\mathbf{r}, t)  \cdot \mathbf{n}\mathrm{d} S$ is over the surface the membrane. In order to simple calculation, we define $\mathcal{M}_{zz}=\left(\int_{\partial V} \stackrel{\leftrightarrow}{\mathbf{M}}(\mathbf{r}, t)  \cdot \mathbf{n}\mathrm{d} S\right)_{z}$. Following the calculation, we can divide $\mathcal{M}_{zz}$ into spin and orbit prats \cite{Barnett_2002},
\begin{equation}
\mathcal{M}_{z z}=\mathcal{M}_{z z}^{\text {spin }}+ \mathcal{M}_{z z}^{\text {orbit }}
\end{equation}
where
\begin{equation}
\mathcal{M}_{z z}^{\mathrm{spin}}=\frac{\epsilon_{0} c^{2}}{2 \omega} \operatorname{Re}\left[-\mathrm{i} \iint \rho \mathrm{d} \rho \mathrm{d} \phi\left(\mathcal{E}_{x} \mathcal{B}_{x}^{*}+\mathcal{E}_{y} \mathcal{B}_{y}^{*}\right)\right]
\end{equation}
and
\begin{equation}
\mathcal{M}_{z z}^{\text {orbit }}=\frac{\epsilon_{0} c^{2}}{4 \omega} \operatorname{Re}\left[-\mathrm{i} \iint \rho \mathrm{d} \rho \mathrm{d} \theta(-\mathcal{B}_{x}^{*} \frac{\partial}{\partial \theta} \mathcal{E}_{y}+\mathcal{E}_{y} \frac{\partial}{\partial \theta} \mathcal{B}_{x}^{*}
-\mathcal{E}_{x} \frac{\partial}{\partial \theta} \mathcal{B}_{y}^{*}+\mathcal{B}_{y}^{*} \frac{\partial}{\partial \theta} \mathcal{E}_{x})\right].
\end{equation}

Under paraxial approximation, the corresponding E.M. fields are
\begin{equation}
\mathcal{E}_{x}=F_{m}(\rho)e^{im\theta}T_{x}(z)\hat{a}_{x},
\quad
\mathcal{E}_{y}=F_{m}(\rho)e^{im\theta}T_{y}(z)\hat{a}_{y},
\end{equation}  
\begin{equation}
\mathcal{B}_{x}=-\dfrac{F_{m}(\rho)e^{im\theta}}{i\omega}\dfrac{\partial{T_{y}(z)}}{\partial{z}}\hat{a}_{y},
\quad
\mathcal{B}_{y}=\dfrac{F_{m}(\rho)e^{im\theta}}{i\omega}\dfrac{\partial{T_{x}(z)}}{\partial{z}}\hat{a}_{x},
\end{equation}  
where $m$ is quantum number of the orbital angular momentum, $F_{m}(\rho)$ is the radial mode function and $T_{x}(z)$, $T_{y}(z)$ are the longitudinal mode functions. 
Without loss of generality, we assume that $\iint d\rho d\phi|F(\rho)|^{2}=1$. It is worth to notice that $\mathcal{B}_{x}$ is related with operator $\hat{a}_{y}$ and $\mathcal{B}_{y}$ is related with operator $\hat{a}_{x}$. Using these equations, we can derive the operator of optical spin angular momentum flux $\mathcal{M}_{zz}^{\text{spin}}$:
\begin{equation}
\begin{aligned}
\mathcal{M}_{z z}^{\mathrm{spin}}&=\frac{\epsilon_{0} c^{2}}{2 \omega} \operatorname{Re}\left[-\mathrm{i} \iint \rho \mathrm{d} \rho \mathrm{d} \theta\left(\mathcal{E}_{x} \mathcal{B}_{x}^{*}+\mathcal{E}_{y} \mathcal{B}_{y}^{*}\right)\right]\\
&=\frac{\epsilon_{0} c^{2}}{4 \omega^{2}} \left[
T_{y}(z)\dfrac{dT_{x}^{*}(z)}{dz}\hat{a}_{x}^{\dagger}\hat{a}_{y}-T_{x}(z)\dfrac{dT_{y}^{*}(z)}{dz}\hat{a}_{y}^{\dagger}\hat{a}_{x}
\right]+ h.c.
\label{spin}
\end{aligned}
\end{equation}
Similarly, the optical orbital angular momentum flux $\mathcal{M}_{zz}^{\text{orbit}}$ is
\begin{equation}
\begin{aligned}
\mathcal{M}_{zz}^{\text{orbit}}&=m\frac{\epsilon_{0} c^{2}}{2 \omega}\operatorname{Re}\left[
\iint\rho d\rho d\theta (-\mathcal{B}_{x}^{*}\mathcal{E}_{y}+\mathcal{E}_{x}\mathcal{B}_{y}^{*})
\right]\\
&=m\frac{\epsilon_{0} c^{2}}{2 \omega^{2}}
\operatorname{Re}\left[i\dfrac{dT_{y}^{*}(z)}{dz}T_{y}(z)\right]\hat{a}_{y}^{\dagger}\hat{a}_{y}+
m\frac{\epsilon_{0} c^{2}}{2 \omega^{2}}
\operatorname{Re}\left[i\dfrac{dT_{x}^{*}(z)}{dz}T_{x}(z)\right]\hat{a}_{x}^{\dagger}\hat{a}_{x}.
\end{aligned}
\label{orbit}
\end{equation}
Notice that spin angular momentum flux $\mathcal{M}_{zz}^{\text{spin}}$ involves the exchange interaction between two modes $\hat{a}_{x}$ and $\hat{a}_{y}$, while orbital angular momentum flux $\mathcal{M}_{zz}^{\text{orbit}}$ is proportional to photonic energy $\hat{a}_{x}^{\dagger}\hat{a}_{x}$ and $\hat{a}_{y}^{\dagger}\hat{a}_{y}$.
If the input beam does not carry orbital angular momentum, which means $m=0$, we can easily prove that $\mathcal{M}_{zz}^{\text{orbit}}=0$ but $\mathcal{M}_{zz}^{\text{spin}}$ is not influenced.
If the input beam is a single frequency E.M. field, we can easily prove that $\tau_{\mathrm{field}}=0$. Then considering the condition $d\ll \lambda$, the mechanical torque $\tau_{z}$ is 
\begin{equation}
\iint dxdy( M_{zz}(x,y,0^{-}) -M_{zz}(x,y,0^{+}) ) =\tau_{\mathrm{mech}}.
\end{equation}
Here we assume the 
dielectric membrane is in $z=0$.
The corresponding angular momentum optomechanical Hamiltonian is
\begin{equation}
H_{\mathrm{OM}}=-\hat{\tau}_{\mathrm{mech}}\hat{\theta}=-\hat{\tau}_{\mathrm{mech}}^{\mathrm{spin}}\hat{\theta}-\hat{\tau}_{\mathrm{mech}}^{\mathrm{orbit}}\hat{\theta}=H_{\text{OM}}^{\text{Spin}}+H_{\text{OM}}^{\text{Orbit}}
\label{OM H},
\end{equation}
where $H_{\text{OM}}^{\text{Spin}}=-\hat{\tau}_{\mathrm{mech}}^{\mathrm{spin}}\hat{\theta}$ and $H_{\text{OM}}^{\text{Orbit}}=-\hat{\tau}_{\mathrm{mech}}^{\mathrm{Orbit}}\hat{\theta}$. Considering Equ.(\ref{spin}) and Equ.(\ref{orbit}), we know that 
\begin{equation}
H_{\text{OM}}^{\text{Spin}} \propto (g_{\text{spin}}\hat{a}_{x}^{\dagger}\hat{a}_{y} + g_{\text{spin}}^{*}\hat{a}_{y}^{\dagger}\hat{a}_{x})\hat{\theta},\quad H_{\text{OM}}^{\text{Orbit}} \propto m( g_{x}\hat{a}_{x}^{\dagger}\hat{a}_{x} +g_{y}\hat{a}_{y}^{\dagger}\hat{a}_{y} )\hat{\theta}.
\end{equation}
Essentially, our optomechanical system can be viewed as the stacking of several optical anisotropic dielectric membranes (Fig. \ref{fig:example}(b)). Therefore, the optomechanical Hamiltonian of our system also has the form $H_{\text{OM}}\propto (g a_{x}^{\dagger}a_{y} + h.c.)\hat{\theta}=(g a_{x}^{\dagger}a_{y} + h.c.)(b+b^{\dagger})$.

\section{Discussion about input-output relation}
\subsection{Mechanical Mode }
When the mechanical frequency $\Omega$ is much smaller than the optical pulse
bandwidth, $\Omega \ll \tau^{-1}$, the mechanical oscillator Hamiltonian $H_{M}=\hbar \Omega b^{\dagger}b$ and mechanical damping characterized by the mechanical damping rate $\gamma=\Omega/Q_{t}$ can be neglect in the time scale of a single optical pulse. 

\subsection{Optical Mode $a_{2}$}
If we use the original form of $H_{OM}$ (\ref{HOM}), we can derive the Langevin equation of the optical mode $a_{2}$,  
\begin{equation}
 \Dot{a_{2}}=-\dfrac{\kappa}{2}a_{2}-\sqrt{2}g \theta_{M}a_{1}+\sqrt{\kappa}a_{2in}(t)   
 \label{a2 exa}
\end{equation}
where $\kappa$ is the decaying rate of the cavity. If the input mode $a_{2}(t)$ is a intense coherent laser pulse with duration $\tau$, we can neglect weak coupling term $\sqrt{2}g\theta_{M}a_{1}$. In our following discussion, we only consider the classical amplitude of mode $a_{2}$ and we define that $\langle a_{2} \rangle=\alpha_{2}$. When $\kappa \gg \tau^{-1}$, we can eliminate the dynamics of the intra-cavity field $a_{2}$ adiabatically, yielding a simple relation between the external and internal field amplitudes:
\begin{equation}
\frac{d \alpha_{2}(t)}{d t}\approx-\dfrac{\kappa}{2} \alpha_{2}(t)+\sqrt{\kappa} \alpha_{2\text{in}}(t) \approx 0 \quad \Rightarrow \quad \alpha_{2}(t)=\dfrac{2}{\sqrt{\kappa}} \alpha_{2\text{in}}(t).
\label{a2}
\end{equation}
We assume the temporal function of $a_{2\text{in}}(t)$ is $f(t)$ with an envelop of duration $\tau$. $f(t)$ satisfies the normalized condition $\int_{-\infty}^{\infty}f(t)^{2}dt=1$.
If the photon number of a single pulse is $N_{\text{in}}$, then $\alpha_{2\text{in}}$ becomes $\alpha_{2\text{in}}(t)=\sqrt{N_{\text{in}}}f(t)$. It is easy to verify that $\int_{-\infty}^{\infty}|\alpha_{2\text{in}}(t)|^{2}dt=N_{\text{in}}$, which shows the photon number of single pulse is $N_{\text{in}}$. At this case, $\alpha_{2}(t)$ can be exprerssed as 
\begin{equation}
	\alpha_{2 }(t)=\dfrac{2}{\sqrt{\kappa}} \alpha_{2\text{in}}(t)=2\sqrt{\dfrac{N_{\text{in}}}{\kappa}}f(t)=\alpha f(t),
	\label{alpha2}
\end{equation}
where $\alpha=2\sqrt{\dfrac{N_{\text{in}}}{\kappa}}$ is the amplitude of $\alpha_{2 }(t)$.

\subsection{Input-Output Relation}
Neglecting phonon energy $H_{\mathrm{m}}=\hbar\Omega b^{\dagger}b$ and using $\alpha_{2}(t)$ to replace operator $a_{2}$, we can describe 
the dynamics of the optical and mechanical quadrature operators by the set
of the following Langevin equations:
\begin{equation}
\begin{aligned}
\frac{d x_{L}(t)}{d t} &=-\dfrac{\kappa}{2} x_{L}(t)+\sqrt{\kappa} x_{L}^{\text{in}}(t), \\
\frac{d p_{L}(t)}{d t} &=-2 g\alpha_{2}(t) \theta_{M}-\dfrac{\kappa}{2} p_{L}(t)+\sqrt{\kappa} p_{L}^{\text{in}}(t) ,\\
\frac{d \theta_{M}}{d t} &=0, \\
\frac{d L_{M}}{d t} &=-2 g\alpha_{2}(t) x_{L}(t)
\end{aligned}
\end{equation}
where $x_{L}^{\text{in}}(t)$ and $p_{L}^{\text{in}}(t)$ are the quadratures of input noise, which satisfy the commutation relation $\left[x_{L}^{\text{in}}(t), p_{L}^{\text{in}}\left(t^{\prime}\right)\right]=i \delta\left(t-t^{\prime}\right)$. 
Invoking adiabatic elimination of the optical intra-cavity field $\frac{d x_{L}(t)}{d t} \approx 0$ and $\frac{d p_{L}(t)}{d t} \approx 0$, we will have the following equations
\begin{equation}
x_{L}(t)=\dfrac{2}{\sqrt{\kappa}}x_{L}^{\text{in}}(t), \quad p_{L}(t)=\dfrac{2}{\sqrt{\kappa}}p_{L}^{\text{in}}(t)-\dfrac{4g\alpha_{2}(t)}{\kappa}\theta_{M}.
\label{ada}
\end{equation}
By means of the usually optical input-output relation $a_{1}^{\text{out}}(t)+a_{1}^{\text{in}}(t)=\sqrt{\kappa} a_{1}(t)$ we find:
\begin{equation}
\begin{aligned}
x_{L}^{\text{out}}(t) &=x_{L}^{\text{in}}(t), \\
p_{L}^{\text{out}}(t) &=p_{L}^{\text{in}}(t)-\frac{4 g \alpha_{2 }(t)}{\sqrt{\kappa}} \theta_{M}(t), \\
\frac{d \theta_{M}}{d t} &=0, \\
\frac{d L_{M}}{d t} &=-2 g\alpha_{2}(t)x_{L}(t),
\label{input 1}
\end{aligned}
\end{equation}

In order to derive the inout-output in our main article, we define time collective optical quandratures  $x_{L}^{\text{in/out}}=\int_{-\infty}^{\infty}f(t)x_{L}^{\text{in/out}}(t)dt$, $p_{L}^{\text{in/out}}=\int_{-\infty}^{\infty}f(t)p_{L}^{\text{in/out}}(t)dt$. It is easy to verify that $x_{L}^{\text{in/out}}$ and $p_{L}^{\text{in/out}}$ also satisfy the standard commutation 
$\left[x_{L}^{\text{in/out}}, p_{L}^{\text{in/out}}\right]=i$. Besides, we define $\theta_{M}^{\text{ in}}=\theta_{M}(t=-\infty)$ and $\theta_{M}^{\text{ out}}=\theta_{M}(t=+\infty)$. As $\alpha_{2 }(t)$ is a short pulse, $\theta_{M}^{\text{ in}}$ and $\theta_{M}^{\text{ out}}$ represent the $\theta_{M}$ quadrature before and after the optomechanical interaction, respectively.

Firstly, from equation $\frac{d \theta_{M}}{d t} =0$ we known that $\theta_{M}(t)=\theta_{M}^{\text{ in}}=\theta_{M}^{\text{ out}}$, which means that $\theta_{M}(t)$ is a constant operator during the optomechanical interaction.
Secondly, multipling the equation $x_{L}^{\text{out}}(t) =x_{L}^{\text{in}}(t)$ by $f(t)$ and integrating it over the whole time interval, we will obtain the equation $x_{L}^{\text {out }} =x_{L}^{\text {in }}$. Similarly, also multipling the equation $p_{L}^{\text{out}}(t) =p_{L}^{\text{in}}(t)-\frac{4 g \alpha_{2 }(t)}{\sqrt{\kappa}} \theta_{M}$ by $f(t)$ and integrating it over the whole time interval, we will have $p_{L}^{\text{out}}=p_{L}^{\text{in}}-\dfrac{4g}{\sqrt{\kappa}}\int_{-\infty}^{\infty}f(t)\alpha_{2 }(t)\theta_{M}(t)dt=p_{L}^{\text{in}}-\dfrac{4g\theta_{M}^{\text{in}}}{\sqrt{\kappa}}\int_{-\infty}^{\infty}f(t)\alpha_{2 }(t)dt$. 
Defining $\chi=\dfrac{4g}{\sqrt{\kappa}}\int_{-\infty}^{\infty}f(t)\alpha_{2 }(t)dt$, and considering Equ. (\ref{alpha2}) $\alpha_{2}(t)=\alpha f(t)$, we have
\begin{equation}
\chi=\dfrac{4g}{\sqrt{\kappa}}\int_{-\infty}^{\infty}f(t)\alpha_{2 }(t)dt=\dfrac{4g}{\sqrt{\kappa}}\int_{-\infty}^{\infty}\alpha f(t)^{2}dt=\dfrac{4g\alpha}{\sqrt{\kappa}}=\dfrac{8g\sqrt{N_{\text{in}}}}{\kappa}=\dfrac{4g}{\sqrt{\kappa}}\sqrt{\int_{-\infty}^{\infty}|\alpha_{2}(t)|^{2}dt}.
\end{equation}
As a consequence, we have $p_{L}^{\text {out }} =p_{L}^{\text {in }}-\chi \theta_{M}^{\text {in }}$.
Thirdly, from equation $\frac{d L_{M}}{d t} =-2 g\alpha_{2}(t)x_{L}(t)$ we know that $L_{M}^{\text{ out}}-L_{M}^{\text{ in}}=-2g\int_{-\infty}^{\infty}\alpha_{2}(t)x_{L}(t)$. Using Equ. (\ref{ada}) $x_{L}(t)=\dfrac{2}{\sqrt{\kappa}}x_{L}^{\text{in}}(t)$, we will derive the equation $L_{M}^{\text{ out}}-L_{M}^{\text{ in}}=-4g/\sqrt{\kappa}\int_{-\infty}^{\infty}\alpha_{2}(t)x_{L}^{\text{in}}(t)dt=-\dfrac{4g\alpha}{\sqrt{\kappa}}x_{L}^{\text{in}}=-\chi x_{L}^{\text{in}}$.

Finally, we will derive this set of input-output relation
\begin{equation}
\begin{aligned}
x_{L}^{\text {out }} &=x_{L}^{\text {in }}, \\
p_{L}^{\text {out }} &=p_{L}^{\text {in }}-\chi \theta_{M}^{\text {in }}, \\
\theta_{M}^{\text {out }} &=\theta_{M}^{\text {in }}, \\
L_{M}^{\text {out }} &=L_{M}^{\text {in }}-\chi x_{L}^{\text {in }}.
\end{aligned}
\label{inout}
\end{equation}

\section{Wigner Function Formaluism}

For a continuous variable quantum system, if the initial states, the system evolution  and the measurement process are all Gaussian, the whole system can advantageously be modeled using standard Gaussian formalism. However, in our system, the 
the input optical state is non-Gaussian state and we must use other method to derive the Wigner function of the resulting mechanical state. 
\subsection{General Definition of Wigner function}
For a $n$ modes Bosonic system, the Wigner function is defined by
\begin{equation}
	W(\boldsymbol{r})=\frac{1}{(2 \pi)^{2 n}} \int \mathrm{d}^{2 n} \boldsymbol{\beta} \operatorname{Tr}\{\hat{\rho} \hat{D}(\boldsymbol{\beta})\} \exp \{-\mathrm{i} \boldsymbol{r}^{T} \cdot \Omega \boldsymbol{\beta}\},
\end{equation}
in which the $2 n \times 2 n$ matrix $\Omega$ is the $n$ -fold block diagonal matrix with diagonal blocks $	\varpi=\left(\begin{array}{rr}
		0 & 1 \\
		-1 & 0
	\end{array}\right)$
and $\hat{D}(\boldsymbol{\beta})$ is the multi-mode displacement operator with $\boldsymbol{\beta}^{\mathrm{T}}=\left(\Re\left\{\beta_{1}\right\}, \Im\left\{\beta_{1}\right\}, \Re\left\{\beta_{2}\right\}, \ldots\right)$ giving the displacement of each mode.
Points in phase space are denoted $\boldsymbol{r}^{\mathrm{T}}=\left(x_{1}, p_{1}, x_{2}, \ldots\right)$ and $\boldsymbol{r} \cdot \Omega \boldsymbol{\beta}=\sum_{i=1}^{n} (-p_{i} \Re\left\{\beta_{i}\right\}+x_{i}\Im\left\{\beta_{i}\right\})$.
For a single mode state, displacement operator $\hat{D}(\boldsymbol{\beta})$ is defined as 
$
\hat{D}(\boldsymbol{\beta})=\exp(\beta a^{\dagger}-\beta a).
$
If we define $\hat{x}=a^{\dagger}+a$, $\hat{p}=i(a^{\dagger}-a)$, and $ \hat{\boldsymbol{r}}=(\hat{x},\hat{p})$, then
$
\beta a^{\dagger}-\beta a=i (\Im\left\{\beta\right\} \hat{x}-\Re\left\{\beta\right\} \hat{p})=i\hat{\boldsymbol{r}}^{T} \cdot \Omega \boldsymbol{\beta}.
$
Therefore, displacement operator can also be expressed as 
$\hat{D}(\boldsymbol{\beta})=\exp(i\hat{\boldsymbol{r}} \cdot \Omega \boldsymbol{\beta})$.
The Wigner function is Fourier dual to the characteristic function $\chi(\boldsymbol{\beta})=\operatorname{Tr}\{\hat{\rho} \hat{D}(\boldsymbol{\beta})\},$
\begin{equation}
\chi(\boldsymbol{\beta})=\int \mathrm{d}^{2 n} \boldsymbol{r} W(\boldsymbol{r}) \exp \{+\mathrm{i} \boldsymbol{r}^{T} \cdot \Omega \boldsymbol{\beta}\}
\end{equation}
and it is easy to verify that $\chi(0)=\operatorname{Tr}[\hat{\rho}]=1$.

\subsection{Time evolution of Wigner function}
If the time evolution of all the quadratures $\hat{r}$ of 
a bosonic system can be described by a linear transformation, which means that 
\begin{equation}
	\hat{r}(T)=M\hat{r}(0),
\end{equation} 
then the Wigner function and characteristic function of this system at time $T$ can also be totally described by matrix $M$ and the original density matrix $\hat{\rho}(0)$. In order to connect the Wigner function $W(\boldsymbol{r},T)$ of this system at time $T$ with the Wigner function $W(\boldsymbol{r},0)$, we consider the following calculation.  
Firstly, we consider the time evolution of characteristic function.
At time $T$, the displacement operator is 
\begin{equation}
	\hat{D}(\boldsymbol{\beta},T)=\exp[i\hat{\boldsymbol{r}}(T)^{T} \cdot \Omega \boldsymbol{\beta}]=\exp[i(M\hat{\boldsymbol{r}}(0))^{T}  \cdot \Omega \boldsymbol{\beta}]=\exp[i\beta^{T}\Omega^{T}M\hat{\boldsymbol{r}}(0)].
\end{equation}
If we define a vector $\gamma=\Omega M^{\mathrm{T}} \Omega \boldsymbol{\beta}$, then we will have the following relation
\begin{equation}
\begin{aligned} \boldsymbol{\beta}^{\mathrm{T}} \Omega^{\mathrm{T}} M \boldsymbol{\hat{r}}(0) &=-\boldsymbol{\beta}^{\mathrm{T}} \Omega^{\mathrm{T}} M \Omega \Omega^{\mathrm{T}} \boldsymbol{\hat{r}}(0) \\ &=\boldsymbol{\beta}^{\mathrm{T}}\left(-\Omega^{\mathrm{T}} M \Omega\right) \Omega^{\mathrm{T}} \boldsymbol{\hat{r}}(0) \\ &=\boldsymbol{\beta}^{\mathrm{T}}(\Omega M \Omega) \Omega^{\mathrm{T}} \boldsymbol{\hat{r}}(0) \\ &=\gamma^{\mathrm{T}} \Omega^{\mathrm{T}} \boldsymbol{\hat{r}}(0)
\\ &=\boldsymbol{\hat{r}}(0)^{T}\cdot \Omega \gamma.
\end{aligned}
\end{equation}
Therefore, at time $T$
\begin{equation}
\hat{D}(\boldsymbol{\beta},T)=\exp{[ \boldsymbol{\hat{r}}(0)^{T}\cdot \Omega \gamma]}=\hat{D}(\boldsymbol{\gamma},0).
\end{equation}
The corresponding characteristic is 
\begin{equation}
	\chi(\boldsymbol{\beta},T)=\operatorname{Tr}[\hat{\rho}(T)\hat{D}(\boldsymbol{\beta},0)]=\operatorname{Tr}[\hat{\rho}(0)\hat{D}(\boldsymbol{\beta},T)]=\chi(\boldsymbol{\gamma,0}).
\end{equation}

Secondly, we consider the time evolution of Wigner function. If we define $\Omega M^{T} \Omega=K$, then $\gamma=K\boldsymbol{\beta}$.
At time $T$,
\begin{equation}
	W(\boldsymbol{r},T)=\frac{1}{(2 \pi)^{2 n}} \int \mathrm{d}^{2 n} \boldsymbol{\beta} \chi(\boldsymbol{\beta},T) \exp \{-\mathrm{i} \boldsymbol{r}^{T} \cdot \Omega \boldsymbol{\beta}\}=\frac{1}{(2 \pi)^{2 n}} \int \mathrm{d}^{2 n} \boldsymbol{\beta} \chi(K\boldsymbol{\beta},0) \exp \{-\mathrm{i} \boldsymbol{r}^{T} \cdot \Omega \boldsymbol{\beta}\}.
\end{equation}
Using coordination transformation $\beta^{\prime}=K\beta$, then
\begin{equation}
	\boldsymbol{r}^{T} \cdot \Omega \beta = \boldsymbol{r}^{T} \cdot \Omega K^{-1} \beta^{\prime}= \boldsymbol{r}^{T} \cdot (M^{-1})^{T}\Omega^{-1}  \beta^{\prime} =(M^{-1}\boldsymbol{r})^{T} \cdot \Omega \beta^{\prime}.
\end{equation}
Therefore, the Wigner function of the total system at time $T$ is 
\begin{equation}
W(\boldsymbol{r},T) \propto \frac{1}{(2 \pi)^{2 n}} \int \mathrm{d}^{2 n} \boldsymbol{\beta}^{\prime} \chi(\boldsymbol{\beta}^{\prime},0) \exp \{-\mathrm{i} (M^{-1}\boldsymbol{r})^{T} \cdot \Omega \boldsymbol{\beta}^{\prime}\} = W(M^{-1}\boldsymbol{r},0).
\label{wig time}
\end{equation}
Here we use $"\propto"$ instead of $"="$ is that we do not calculate the Jacobin matrix during the process of coordination transformation $\beta^{\prime}=K\beta$.

\subsection{The Resulting Wigner Function  }
For our optomechanical system, if we neglect all of the noises and only care about the torsional mode $b$ and optical mode $a_{1}$, the transformation matrix $M$ is
\begin{equation}
	M=\begin{pmatrix}
	1& 0 & 0 &0  \\
	0&  1& -\chi &0  \\
	0& 0 & 1 & 0 \\
	-\chi& 0 & 0 & 1
	\end{pmatrix},
\end{equation}
which is derived from the input-output relation ($\ref{inout}$).

We assume the initial quantum state $\rho(0)=\rho_{L}\otimes\rho_{M}$, where $\rho_{L}$ is the density matrix of the input optical mode $a_{1\text{in}}$ and $\rho_{M}$ is the density matrix of the mechanical mode. The initial Wigner function is 
$W(\boldsymbol{r},0)=W_{L}(x_{L},p_{L})W_{M}(x_{M},p_{M})$. After the optomechanical interaction ($\ref{inout}$), the Wigner function of this system is 
\begin{equation}
    W(\boldsymbol{r},T)=W(M^{-1}\boldsymbol{r},0)=W_{L}(x_{L},p_{L}+\chi x_{M})W_{M}(x_{M},p_{M}+\chi x_{L}).
\end{equation}
If we do a quantum measurement on the subsystem $\rho_{L}$ with POVM measurement operator $\Pi$, then the resulting Wigner function is 
\begin{equation}
	W_{\text{meas}}(x_{M},p_{M}) \propto \int \mathrm{d}^{2} \boldsymbol{r}_{L}W(\boldsymbol{r})W_{\Pi}(x_{L},p_{L}),
	\label{meas}
\end{equation}
where
\begin{equation}
	W_{\Pi}(\boldsymbol{r})=\frac{1}{(2 \pi)^{2 n}} \int \mathrm{d}^{2 n} \boldsymbol{\beta} \operatorname{Tr}\{\Pi \hat{D}(\boldsymbol{\beta})\} \exp \{-\mathrm{i} \boldsymbol{r}^{T} \cdot \Omega \boldsymbol{\beta}\},
\end{equation}
which is the Wigner function of the measurement operator $\Pi$. In the field of quantum optics, the most commonly used measurements are photon number detection, homodyne detection and heterodyne detection.
\paragraph{Photon Number Detection}
The POVM operator for photon number resolved detection with detection efficiency $\eta$
is $\Pi=\Pi_{m}(\eta)=\eta^{m} \sum_{k=m}^{\infty}(1-\eta)^{k-m}
C_{m}^{k}
|k\rangle\langle k|$ where $C_{m}^{k}=\dfrac{m!}{k!(m-k)!}$. The corresponding Wigner function \cite{0503237}
\begin{equation}
W(X,Y)=\frac{1}{2\pi} \frac{(-1)^{m} \eta^{m}}{(2-\eta)^{1+m}} L_{m}\left(\frac{(X^2+Y^2)}{2-\eta}\right) \exp \left\{-\frac{ \eta}{2(2-\eta)}(X^2+Y^2)\right\}
\label{pho}
\end{equation}
where $L_{k}(x)$ is a Laguerre polynomials.

\begin{figure}
	\centering
	\includegraphics[width=1\linewidth]{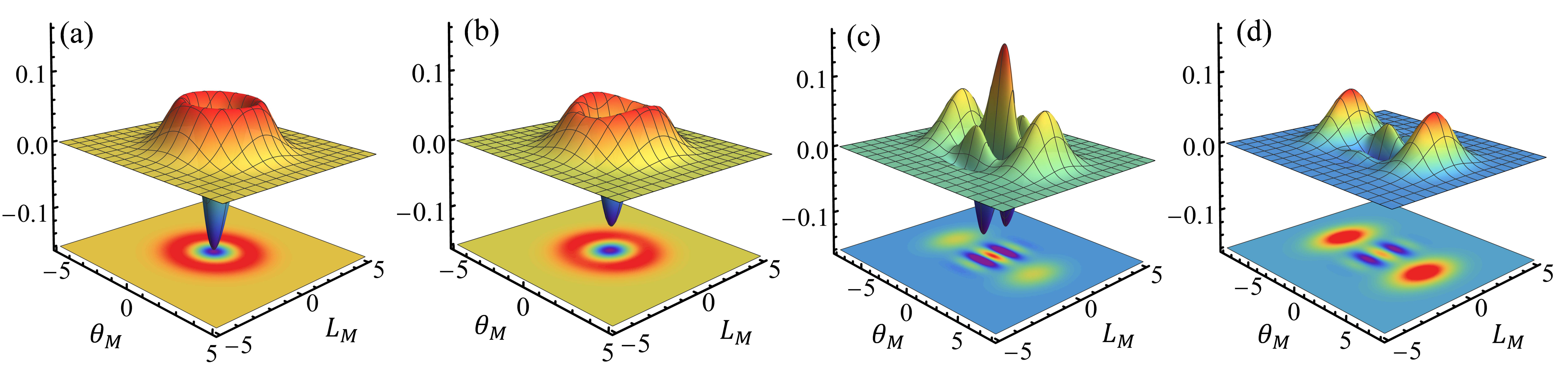}
	\caption{(a) Wigner Function of single phonon fock state. (b) Wigner function $W_{\text{sfcok}}(\theta_{M},L_{M})$. Here we set $V_{LL}=2000$ and $V_{\theta\theta}=0.2$. (c) Wigner function of cat state with $\alpha=2$. (d)  Wigner function $W_{\text{scat}}(\theta_{M},L_{M})$. Here we also set $V_{LL}=2000$ and $V_{\theta\theta}=0.2$. }
	\label{fig:com}
\end{figure}

\paragraph{Homodyne Detection}
If we use a photodetector with efficiency $\eta=1$ to detect the quadrature operator $X_{\phi}=a e^{-i\phi}+ a^{\dagger} e^{i\phi}$ with the measurement result $x$, then the corresponding POVM is 
$
\Pi (x)=|x \rangle_{\phi} \langle x|
$,
where $X_{\phi} | x\rangle_{\phi}=x| x\rangle_{\phi}$
and the corresponding Wigner functions is given \cite{0503237}
\begin{equation}
W(X,P) \propto \delta\left(x-X\cos(\phi)-Y\sin(\phi)\right).
\end{equation}

In our article, after the ideal homodyne detection of the output optical mode, the Wigner function of the resulting mechanical state is 
\begin{equation}
	W_{\text{meas}}(\theta_{M},L_{M})\propto\int \mathrm{d}^{2}\boldsymbol{r}_{L} W_{L}(x_{L},p_{L}+\chi \theta_{M})W_{M}(\theta_{M},L_{M}+\chi x_{L})\delta(p_{L}-p)
	\label{general w}
\end{equation}
where $p$ is the result of the homodyne detection and in our numerical calculation we will choose $p=0$.
\paragraph{Single Photon Pulse Input}
Here we consider a special and simple example: the original input optical state is a single photon fock state pulse with Wigner function $W_{L}(x,p)\propto \exp{(-\dfrac{1}{2}x^{2}-\dfrac{1}{2}p^{2})}(x^{2}+p^{2}-1)$ 
and the original mechanical state is a precooled thermal squeezed state $W_{M}(\theta_{M},L_{M})\propto \exp\left[ -\dfrac{\theta_{M}^{2}}{2V_{\theta \theta}} -\dfrac{L^{2}}{2V_{LL}} \right]$
where $V_{LL}$ and $V_{\theta\theta}$ are the vatiance of $L_{M}$ and $\theta_{M}$, respectively.
When $p=0$, the resulting mechanical state is
\begin{equation}
W_{\text{sfock}}(\theta_{M},L_{M}) \propto \exp{[-\dfrac{1}{2}(1+V_{LL}^{-1})\theta_{M}^{2}-\dfrac{1}{2}(1+V_{\theta\theta})^{-1}L_{M}^{2}]} 
\left[(1+V_{\theta\theta})\theta_{M}^2+(1+V_{\theta\theta})^{-1}L_{M}^2-1
\right].
\end{equation}
If $V_{LL}\gg 1$, Wigner function $W_{\text{sfock}}(\theta_{M},L_{M})$ will near the Wigner function of squeezed single phonon fock state and 
 when $V_{LL} \to \infty$ and $V_{\theta\theta}\to 0$,
$W_{\text{meas}}(\theta_{M},L_{M})$ will near the Wigner function of single phonon fock state (see Fig. \ref{fig:com}).

\paragraph{Perfect Cat State Input}
Now we consider a more complex case: the input optical state is a perfect 
even cat state $|Cat\rangle \propto |\alpha\rangle + |-\alpha\rangle$. Without loss of generality, we assume that $\alpha$ is positive real value. For this even cat state, the corresponding Wigner function is 
\begin{equation}
W_{\alpha}(\boldsymbol{r})=\frac{\mathrm{e}^{-\frac{1}{2} r \cdot r}}{2 \pi\left(1 + \mathrm{e}^{-2|\alpha|^{2}}\right)}\left[\mathrm{e}^{-2|\alpha|^{2}} \cosh (2 \boldsymbol{r} \cdot \boldsymbol{\alpha}) + \cos (2 \boldsymbol{r} \cdot \varpi \boldsymbol{\alpha})\right],
\end{equation}
where $\boldsymbol{r}=(x_{L},p_{L})$ and  $	\varpi=\left(\begin{array}{rr}
0 & 1 \\
-1 & 0
\end{array}\right)$. As for the mechanical state $W_{M}(\theta_{M},L_{M})$, we also assume that it is a thermal squeezed state.
Assuming the homodyne result $p=0$, the Wigner function of the resulting mechanical state is
\begin{equation}
    W_{\text{scat}}(\theta_{M},L_{M}) \propto \exp{[-\dfrac{1}{2}(1+V_{LL}^{-1})
    	\theta_{M}^{2}-\dfrac{1}{2}(1+V_{\theta\theta})^{-1}L_{M}^{2}]} 
    \left[ 
    \cos(\dfrac{2\alpha}{1+V_{\theta\theta}} L_{M}) + \exp{(-\dfrac{2\alpha^{2}}{1+V_{\theta\theta}})} \cosh(2\alpha\theta_{M})
    \right].
\end{equation}
Notice that  
when $V_{LL} \to \infty$ and $V_{\theta\theta}\to 0$,
$W_{\text{scat}}(\theta_{M},L_{M})$ will near the Wigner function of cat state (see Fig. \ref{fig:com}).

\begin{figure}[ht]
	\centering
	\includegraphics[width=0.85\linewidth]{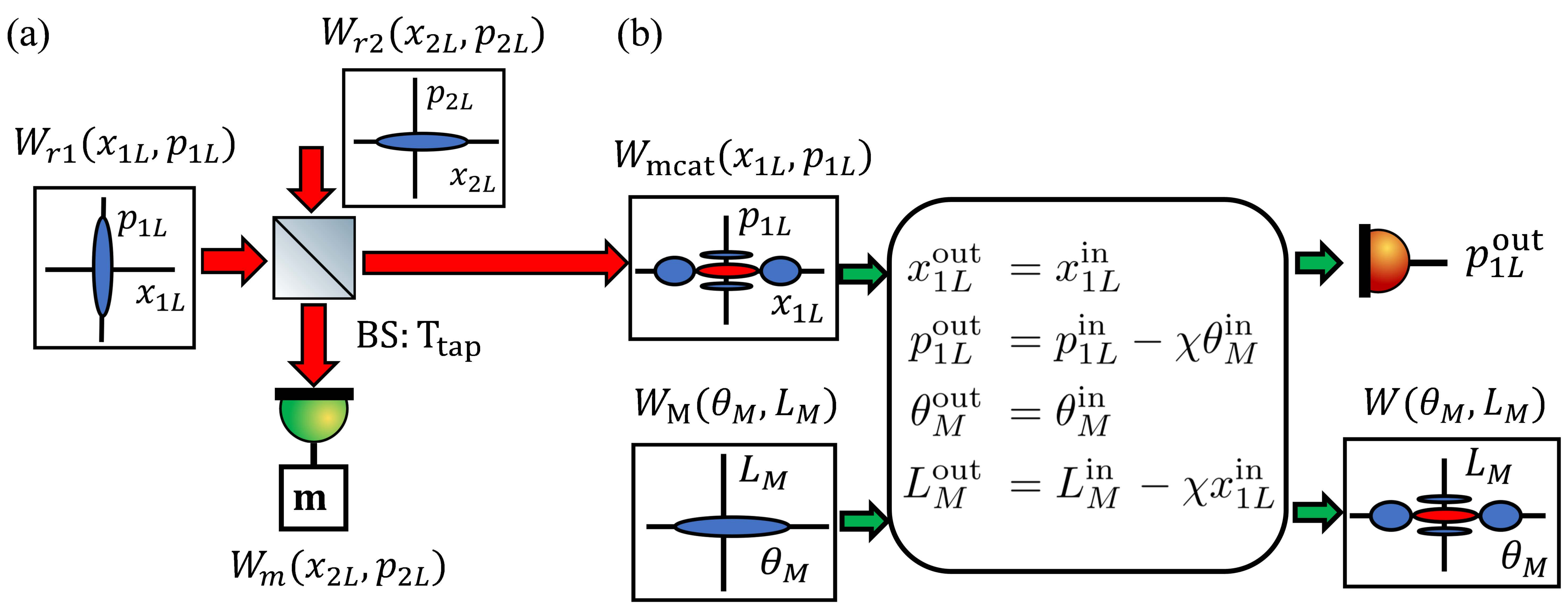}
	\caption{(a) The preparing process of the optical cat state with Wigner function $W_{\text{mcat}}(x_{1L},p_{1L})$. BS is a beam splitter with transmittance $T_{\text{tap}}$, $m$ represents the photon number resolving detection with $m$ photons detected and $W_{m}(x_{2L},p_{2L})$ is the corresponding Wigner function. (b) State transformation and preparing process of the resulting mechanical cat-like state with Wigner function $W(\theta_{M},L_{M})$. The input mechanical state $W_{M}(\theta_{M},L_{M})$ is a squeezed thermal state along $\hat{L}_{M}$ quadrature and $p_{1L}^{\text{out}}$ represents the homodyne detection on the output optical quadrature $\hat{p}_{1L}^{\text{out}}$. }
	\label{fig:process}
\end{figure}

\begin{figure}
	\centering
	\includegraphics[width=1\linewidth]{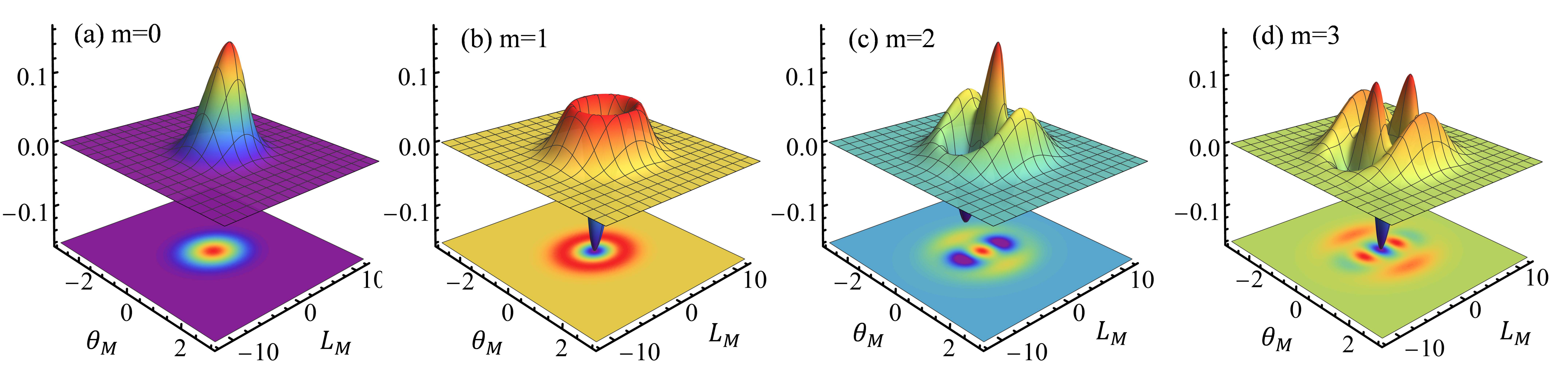}
	\caption{Wigner functions of optical catlike state prepared by generalized photon subtraction method. In numerical calculation, we choose $r_{1}=-r_{2}=1.15$ and $T_{\text{tap}}=(e^{2r_{1}}-1)/(e^{2r_{1}}-e^{-2r_{1}})\approx0.909$. Here $m$ represents the detected photon number.}
	\label{fig:optical-cat}
\end{figure}

\paragraph{Catlike State Prepared by Generalized Photon Subtracted Method}
Following the calculation in ref \cite{PhysRevA.103.013710}, we discuss the Wigner function of the catlike state prepared by generalized photon subtracted method (see Fig. \ref{fig:process}). For a squeezed state with sqeezing parameter $r$ ($r$ is real number), the corresponding Wigner function $W_{r}(x_{L},p_{L})$ is 
\begin{equation}
	W_{r}(x_{L},p_{L}) \propto \exp \left[-\dfrac{x_{L}^{2}}{2\exp(2r)}-\dfrac{p_{L}^{2}}{2\exp(-2r)}\right].
\end{equation} 
In the generalized photon subtracted method, the original optical input state are two unentangled squeezed with squeezing parameter $r_{1}$ and $r_{2}$. The corresponding Wigner function of the input state is 
\begin{equation}
W_{\text{Lin}}(x_{1L},p_{1L},x_{2L},p_{2L}) \propto W_{r_{1}}(x_{1L},p_{1L})W_{r_{2}}(x_{2L},p_{2L}).
\end{equation} 
The asymmetric beam splitter with transmtance $T_{tap}$ exchanges the quantum state of the two squeezed light. If we define $\hat{r}_{L}=(\hat{x}_{1L},\hat{p}_{1L},\hat{x}_{2L},\hat{p}_{2L})$, then after the interaction of the beam splitter, the resulting optical quadratures will be 
\begin{equation}
	\hat{r}_{L} \rightarrow M_{L}\hat{r}_{L},
\end{equation} 
where $M_{L}$ is the transmission matrix of the beam splitter, and the exact form of $M_{L}$ is
\begin{equation}
M_{L}=
	\begin{pmatrix}
	\sqrt{T_{tap}} & 0 & \sqrt{1-T_{tap}} & 0 \\
	0 & \sqrt{T_{tap}} & 0 &\sqrt{1-T_{tap}}  \\
-\sqrt{1-T_{tap}}	& 0 &\sqrt{T_{tap}}  & 0 \\
	0 & -\sqrt{1-T_{tap}} & 0 & \sqrt{T_{tap}}
	\end{pmatrix}.
\end{equation}
According to the time evolution formalism of Wigner function ($\ref{wig time}$), after the beam splitter, the corresponding optical Wigner function $W_{Ltap}$ is 
\begin{equation}
	W_{\text{Ltap}}(x_{1L},p_{1L},x_{2L},p_{2L}) \propto W_{\text{Lin}}(M_{L}^{-1}r_{L}).
\end{equation} 
Then we will do photon number resolved detection on the output port of squeezed light $r_{2}$. The corresponding measurement Wigner function $W_{\text{ m}}(x_{L2},p_{L2})$ is Equ. (\ref{pho}), notice that $W_{\text{m}}(x_{L2},p_{L2})$
depends on the detected photon number $m$. Following the photon number resolved detection, the photon on the other port will be projected into a catlike state, and the exact expression of the resulting Wigner function is 
\begin{equation}
 W_{\text{mcat}}(x_{1L},p_{1L}) \propto \int dx_{2L}dp_{2L}	W_{\text{Ltap}}(x_{1L},p_{1L},x_{2L},p_{2L}) W_{\text{m}}(x_{L2},p_{L2}).
\end{equation}
Following these processes, we prapare a opticl catlike state. The Wigner function of the prepared optical catlike states are plotted in Fig. (\ref{fig:optical-cat}). When $m>1$, the resulting Wigner function shows the feature of catlike state. 

By using this optical catlike sate pulse to pump the optical mode $a_{1}$ in our optomechanical system, we cat prepare our torsional oscillator in a catlike state, and the Wigner function of the resulting mechanical state is 
 
\begin{equation}
W(\theta_{M},L_{M})\propto\int \mathrm{d}^{2}dx_{1L}dp_{1L} W_{\text{mcat}}(x_{1L},p_{1L}+\chi \theta_{M})W_{M}(\theta_{M},L_{M}+\chi x_{L})\delta(p_{1L}-p).
\end{equation}
For different detected photon number $m$, $W_{\text{mcat}}$ will be different, thus $W(\theta_{M},L_{M})$ will also be different.

\bibliographystyle{unsrt}
\bibliography{reference}